\documentclass[letterpaper,dvipsnames]{article}
\pdfoutput=1
\usepackage{jheppub}
\usepackage{amsmath}
\usepackage{tensor}
\usepackage{graphicx}
\usepackage{array}
\usepackage{ulem}
\usepackage{cancel}
\usepackage{xcolor}
\usepackage{slashed}
\usepackage{afterpage}
\usepackage{appendix}
\usepackage{multirow}
\usepackage{float}
\usepackage{mathtools}
\mathtoolsset{centercolon}

\usepackage{multirow}

\usepackage{comment}
\usepackage{multirow}

\newcommand{\aUV}{\alpha_{\rm uv}}
\newcommand{\aIR}{\alpha_{\rm ir}}
\newcommand{\ew}{{\rm ew}}
\newcommand{\vUV}{v_{\rm uv}}
\newcommand{\vIR}{v_{\rm ir}}
\newcommand{\eps}{\epsilon}
\newcommand{\epsb}{{\bar \epsilon}}
\newcommand{\rc}{r_c}
\newcommand{\ir}{{\rm ir}}
\newcommand{\uv}{{\rm uv}}
\newcommand{\eff}{{\rm eff}}

\title{More Effective RS Field Theory}

\author[a]{Severin L\"ust,}
\author[b]{Michael Nee,}
\author[b]{and Lisa Randall}

\affiliation[a]{Sorbonne Université, CNRS, Laboratoire de Physique Théorique et Hautes Énergies, LPTHE, F-75005 Paris, France}
\affiliation[b]{Department of Physics, Harvard University, Cambridge, MA, 02138, USA}
\emailAdd{mnee@fas.harvard.edu}
\emailAdd{randall@g.harvard.edu}

\abstract{In this paper we derive the effective theory for a stabilized five-dimensional warped geometry, addressing several outstanding issues in this derivation. These include allowing for a non-zero 4d cosmological constant, accounting for constraints from both the UV and IR branes, and determining how the stabilized theory responds to an energy perturbation on each brane. We show how a consistent low-energy effective theory from a stable warped solution must respect a constraint that follows from the higher-dimensional Einstein equation. Satisfying the constraint requires that the 4d cosmological constant must be the same everywhere throughout the bulk, which means the stabilizing fields must adjust to allow for the same 4d curvature everywhere in the extra dimension. We show explicitly how this works in a 5d model, and how the correct 4d effective potential reproduces this behavior. We find that the cosmological constant generated from adding energy to one of the branes is unaffected by the details of the stabilization mechanism at leading order, despite the need for the stabilizing fields to adjust. In anticipation of a companion paper, we also  briefly  discuss how supersymmetry can be realized consistently in the 5d theory. In particular, we show how the stabilization mechanism remains consistent with sequestering, even as the supersymmetry-breaking energy is reflected in the stabilizing field throughout the bulk.} 

\begin{document}

\maketitle


\section{Introduction}

Effective theories (EFTs) are ubiquitous within physics, allowing us to evaluate dynamics and potentials purely in the context of low-energy theories.  A carefully constructed EFT, which takes into account all relevant light degrees of freedom and their interactions, allows one to calculate physical quantities up to uncertainties suppressed by the cutoff of the EFT~\cite{Wilson:1973jj, Appelquist:1974tg}. In this paper we show how to construct a consistent effective theory for theories with stabilized extra dimensions while allowing for a nonzero cosmological constant (c.c.) in the 4d theory. Among other things, this has relevance for the communication of supersymmetry breaking and also allows us to address issues with proposed solutions to the cosmological constant problem~\cite{Bellazzini:2013fga, Coradeschi:2013gda, Agrawal:2016ubh}. Studying extra dimensional models with 4d de-Sitter metrics is also relevant for determining the 4d effective theory that describes inflation in these theories, as was considered for example in refs.~\cite{Binetruy:1999hy, Lukas:1999yn, Csaki:1999mp,  Deffayet:2000uy, Kumar:2018jxz, Antoniadis:2023sya, Hubisz:2024xnj}. There is also a large body of literature~\cite{Creminelli:2001th, Nardini:2007me, Hassanain:2007js, Konstandin:2010cd, Konstandin:2011dr, vonHarling:2017yew, Dillon:2017ctw, Bruggisser:2018mrt, Baratella:2018pxi, Megias:2018sxv, Pomarol:2019aae, Agashe:2019lhy, Fujikura:2019oyi, Megias:2020vek, Bigazzi:2020phm, Agashe:2020lfz, Agrawal:2021alq, Nee:2022pxx, Csaki:2023pwy, Girmohanta:2023sjv, Eroncel:2023uqf, Mishra:2023kiu, Mishra:2024ehr, Agrawal:2025wvf, Gherghetta:2025krk} devoted to studying the supercooled phase of the Randall Sundrum (RS) model~\cite{Randall:1999vf, Randall:1999ee} and the constraints that arise from it~\cite{Creminelli:2001th}, which necessarily involves studying the model in 4d de-Sitter space.

In this work we focus on a five-dimensional AdS geometry bounded by two branes~\cite{Randall:1999vf, Randall:1999ee}, with a warped product metric and a static extra dimension, although some of our conclusions are more general. This requires that every point $\phi$ in the 5th dimension has the same 4d metric up to an overall warp factor.  This implies that energy density on every constant-$\phi$ slice must be the same up to two powers of the warp factor, even if the energy density is added locally (for instance on either of the two boundaries). For theories in which the 5d bulk has a background AdS$_5$ metric, the space can be parameterized as a warped version of a 4d metric for which the  4d cosmological constant can take any value that is consistent with the boundary conditions. The lower dimensional theory can be either AdS, dS or Minkowski space~\cite{Karch:2000ct}, but the tension of each of the end-of-the-world branes
must be tuned to the bulk energy density in order for the solution to be static~\cite{Karch:2020iit}. This allows for a derivation of the 4d EFT, in which the time-dependence of the radion responds to any subsequent detuning with energy below the KK scale. This EFT includes the dynamics of the low-energy fields.

Including a stabilization mechanism, such as the Goldberger-Wise (GW) model~\cite{Goldberger:1999uk, Goldberger:1999un} that we consider, alleviates this tuning and allows for a static solution to be found for a range of brane tensions. The main question we answer is how consistent slicing can be maintained after changing the tension of either one of the two branes. Such an energy density changes the 4d cosmological constant, leading to the question of how this energy density is distributed throughout the higher-dimensional space so that each 4d slice has the same metric once we include the perturbation. We show how the warp factor and the stabilizing field, $\Phi$, adjust to satisfy the jump conditions on each brane with the perturbation, allowing for a static extra dimension with the perturbed energy density. These adjustments are seen only when including the backreaction on the metric from $\Phi$ and allowing the boundary values of $\Phi$ to adjust. Our work extends what was done in ref.'s~\cite{Binetruy:1999hy, Csaki:1999mp, Gubser:1999vj, Csaki:2000zn, Maartens:2003tw}, which constructed general cosmological solutions in the RS model, but did not consider in detail how the stabilization mechanism adjusts to allow for a consistent EFT. 

One context in which this study is of interest is the KKLT model~\cite{Kachru:2003aw}, since perturbing the energy density on the IR brane mimics the  addition of antibrane energy in the IR to what is assumed to be a consistent geometry with negative 4d cosmological constant.  What is unclear is how the energy density, which breaks supersymmetry, is communicated to the UV brane, which we know to be essential since both boundary conditions need to be consistent with the perturbed energy density, even though the perturbation is explicit only in the IR.

Another motivation for this study is to better understand how the AdS/CFT correspondence~\cite{Maldacena:1997re, Witten:1998qj, Gubser:1998bc} can be maintained in the presence of perturbations away from pure AdS space.  This study also helps us understand how the Goldberger-Wise mechanism, which specifies UV and IR boundary conditions, can be consistent with a holographic formulation in which boundary conditions are specified only in the UV. In the CFT picture~\cite{Rattazzi:2000hs, Arkani-Hamed:2000ijo}, the shift in $\Phi$ in response to a change in energy in the IR can be thought of as a modified UV boundary condition for $\Phi$. This adjusts the UV theory so that the RG flow of the CFT goes to an IR theory with extra energy density. The requirement of a consistent slicing implies that the cosmological constant is the same at each RG scale in the CFT. This may have relevance to the cosmological energy density, since it means that a consistent 4d theory requires the UV theory to adjust to reflect IR energy.

We also show that the adjustment of the stabilizing fields does not spoil sequestering. Although the stabilization mechanism must adjust to communicate the cosmological constant throughout the bulk, the theory still remains sequestered as the supersymmetry-breaking terms are subleading to the irreducible contribution from gravity via the compensator $F$-term. This is consistent with what has been found in the supersymmetric models considered in ref.'s~\cite{Luty:1999cz, Luty:2000ec, Son:2008mk}.
\\

Deriving the EFT for warped geometry is an old subject and one might think all issues have been resolved. To motivate the rest of the paper, we summarize our principle results:
\begin{itemize}
    \item The bulk of our paper is devoted to showing how the stabilizing field adjusts to accommodate a nonuniform distribution of energy consistent with a fixed 4d cosmological constant, $\Lambda$. We show how $\Lambda$ responds to a change in energy density $\delta T$ on either of the UV and IR branes and show at leading order in $\delta T$ it follows the naive expectation:
    \begin{align}
        \Lambda = \frac{3k}{\kappa^2} \delta T_{\uv / \ir} e^{-4 k \pi r_{\uv / \ir}} \, .
    \end{align}
 
    Our analysis furthermore highlights how the GW stabilization mechanism allows for a solution with the same 4d c.c.\ everywhere in the bulk. 
    
    \item We demonstrate our result for how the energy adjusts both in the full 5d theory and in the effective potential that we obtain by integrating over the extra dimension, and show they agree. Working in the 5d picture requires including the backreaction  from the stabilizing field and 4d cosmological constant on the warp factor, which we treat in a perturbative expansion. The boundary conditions satisfied by the warp factor on each brane determine the cosmological constant.
    
    We repeat our analysis in the 4d effective theory by integrating the action over the extra dimension. The effective potential reduces to a sum of boundary terms plus a contribution from the Einstein-Hilbert term. We will see that the boundary terms are proportional to the boundary conditions, so vanish at the minimum. This shows that keeping the contribution from the Einstein-Hilbert term is necessary to find the correct vacuum energy from the potential in the 4d theory.
    
    \item We allow for general boundary potentials for $\Phi$, generalizing the original model of Goldberger \& Wise. We discuss which combinations of potentials and sign of the $\Phi$ bulk mass can lead to a stabilized model.\footnote{Some of these results have been presented elsewhere in the literature, see e.g.~\cite{Chacko:2013dra, Chacko:2014pqa}.}
    
    \item Having determined how the system responds to a change in energy density on the IR brane, we  comment on how the electroweak phase transition could in principle lead to a large shift in the stabilization radius. This can occur in models where the Higgs is localized on the IR brane and scale invariance is only weakly broken, leading to a radion that is light relative to the electroweak scale. 

    \item Our analysis allows us to address the question: In a supersymmetric model with supersymmetry broken on one brane, how does the energy redistribute so that the same cosmological constant appears on every slice while remaining consistent with sequestering? One might assume that because the stabilization mechanism had to adjust to transfer energy throughout the space that there should be a large change in energy everywhere, thus breaking supersymmetry everywhere and obviating sequestering. If this were true it would have significant implications for anomaly-mediated supersymmetry breaking, which is most relevant when the boundaries of the extra dimension are decoupled~\cite{Randall:1998uk, Giudice:1998xp}. In this work we give general arguments that the supersymmetry breaking communicated by a change in the stabilizing field is subleading to the anomaly-mediated contribution, leaving a more thorough analysis in a supersymmetric model to a companion paper~\cite{Nee:2025nmi}.
\end{itemize}

Previous papers have considered the potential implications of warped geometry for the cosmological constant in scalar gravity theories.
The effective theories that have been derived were sufficient to derive the correct radion potential, but not the cosmological constant. Some authors presented what superficially appear to be very general arguments about the cosmological constant in the low-energy effective theory. We argue that neglecting the UV contribution allows for interesting models that do in fact lead to a suppressed c.c. However, these models fail to capture the essential source of the c.c. in the full theory, which requires including the UV boundary condition. 
\\

The rest of this paper is organized as follows. In section~\ref{sec:slicing} we  set up the 5d theory and show how the assumption of a static extra dimension requires the same 4d metric on every $y$-slice. We also present the warp factor solutions neglecting backreaction from the stabilizing fields. In section~\ref{sec:5dmodel} we add a stabilization mechanism and solve the equations of motion including the leading order backreaction effects. We discuss which choices of boundary potentials lead to stable solutions, and  show how adding a constant energy density on either brane leads to a redistribution of field profiles consistent with a change in the 4d cosmological constant which is the same throughout the bulk. In section~\ref{sec:effectivepotential} we derive the low-energy effective theory including dependence on the stabilizing mechanism. We then show how the effective potential reproduces the 5d results of section~\ref{sec:5dmodel} and illustrate how the cosmological constant we derive is unsuppressed by the parameters of the stabilization mechanism. We briefly discuss what our results imply for supersymmetry breaking and sequestering in section~\ref{sec:SUSY}.

\section{Constraint on Warped Potentials in Maximally Symmetric Slicings} 
\label{sec:slicing}

In this work we are interested in determining the effective theory that comes from the dimensional reduction of models with a compact extra dimension. We focus on 5d models with metrics given by the warped product:
\begin{align}\label{eq:metricansatz}
    ds_5^2 = g^{(5)}_{AB} dx^A dx^B = e^{2 A(\phi)} g_{\mu\nu} dx^\mu dx^\nu +  r_c^2 d\phi^2 \,,
\end{align}
where the spacetime is cut off by two branes located at the endpoints of the space, $\phi = 0, \, \pi$ and $g_{\mu\nu}$ is the 4d metric. In most of the examples we consider, $g_{\mu\nu}$ will either be Minkowski or a  4d de-Sitter metric corresponding to a cosmological constant $\Lambda$, but our results can be extended to a general FRW metric on 4d slices. The size of the extra dimension, $r_c$, is set by a stabilization mechanism that generates a potential for $r_c$. The action we consider is:
\begin{equation}\label{eq:Daction}
    S_5 = \int d^5 x \sqrt{- g^{(5)}} \Bigl( \frac{1}{ 2 \kappa^2} R^{(5)} + \mathcal{L}_\text{mat} \Bigr)  - \sum_i \int\limits_{B_i} d^4x \sqrt{-g_i} \lambda_i \,,
\end{equation}
where $B_i$ is the world-volume for a  brane, $g_i$ is the induced metric on the brane, and $\alpha$ labels the brane. $\kappa$ is the five-dimensional gravitational constant  defined by $\kappa^{-2} = 2 M_5^3$.

The warped product form of the metric~\eqref{eq:metricansatz} assumes that each constant-$\phi$ slice is described by the same 4d metric. For this slicing to be consistent, there must be no time dependence in the $\phi$ direction, so that every position in the fifth dimension has the same 4d cosmological constant~\cite{Karch:2020iit}.\footnote{The effective theory we derive allows for time-dependence when this constraint is not yet satisfied. In the effective theory this corresponds to a time-dependent radion.} It is not obvious {\it a priori} how this is achieved as the vacuum energy could be localized to either of the branes, for example. However, as we will show in later sections, the stabilization mechanism that fixes the 5d radius will also equilibrate the 4d cosmological constant so that the energy density on constant-$\phi$ surfaces is related by the factor $e^{2A(\phi)}$. 
When the destabilizing energy is below the KK scale, the response of the radion and hence the leading time-dependence is captured too.

As pointed out in refs.~\cite{Giddings:2005ff,Douglas:2009zn}, the higher-dimensional Einstein equations should be interpreted as constraints on the 4d effective potential, something which is not derivable from within the effective theory itself. Here we will see that satisfying the higher-dimensional Einstein equations with the warped metric in equation~\eqref{eq:metricansatz} yields the constraint of consistent slicing. The constraint equation is derived as a consistency requirement between the four-dimensional and the five-dimensional Einstein equations~\cite{Douglas:2009zn}, assuming the gradients of fields in the 4d directions vanish. To do so we start with the five-dimensional Einstein equations
\begin{equation}\label{eq:einsteinD}
    R^{(5)}_{MN} - \frac12 g_{MN} R^{(5)} = \kappa^2 T^{(5)}_{MN} \,,
\end{equation}
For the metric~\eqref{eq:metricansatz}, the components of the 5d Ricci tensor $R^{(5)}_{MN}$ and the Ricci scalar $R^{(5)}$ can be written in terms of the warp factor and their 4d counterparts $R_{\mu \nu}$ and $R$ as:
\begin{align} 
    R^{(5)}_{\mu \nu} &= R_{\mu\nu}[g] - {g}_{\mu\nu} \frac{e^{2A}}{r_c^2} \left( A'' + 4 (A')^2 \right) \,, \nonumber \\
    R^{(5)}_{\phi \phi} &=  -4 A'' - 4 (A')^2 \,,
    \label{eq:RicciD}
    \\
    R^{(5)} &= e^{-2A} R  - \frac{4}{r_c^2} \left(2 A'' +5 (A')^2\right) \, .
    \nonumber
\end{align}
Assuming that all 4d derivatives of the fields in $\mathcal{L}_\mathrm{mat}$ vanish, we have for the $tt$-component of the right-hand side of \eqref{eq:einsteinD}:
\begin{equation}
    T^{(5)}_{tt} = e^{2A} g_{tt} \, \mathcal{L}_\text{mat} \,.
    \label{eq:stress_tt}
\end{equation}
For the left hand-side we use 
\eqref{eq:RicciD} to compute their four-dimensional components
\begin{equation}\label{eq:einstein54dpart}
   R^{(5)}_{\mu\nu} - \frac12 g^{(5)}_{\mu\nu} R^{(5)} = R_{\mu\nu} - \frac12 g_{\mu\nu} R + g_{\mu\nu} \frac{e^{2A}}{r_c^2} \left(3 A'' + 6 (A')^2 \right)  \,,
\end{equation}
Using the $tt$ component of the four-dimensional Einstein equations, where $\rho$ is the 4d energy density
\begin{equation}
    \label{eq:4deinstein}
    R_{tt} - \frac12 g_{tt} R = - \frac{\rho}{M_P^2} g_{tt} \,,
\end{equation}
together with~\eqref{eq:stress_tt}, equation~\eqref{eq:einstein54dpart} becomes
\begin{equation}
    R^{(5)}_{tt} - \frac12 g^{(5)}_{tt} R^{(5)} = g_{tt} \left[\frac{e^{2A}}{r_c^2} \left(3 A'' + 6 (A')^2  \right) - \frac{\rho}{M_P^2} \right] \,.
\end{equation}
Inserting everything back into \eqref{eq:einsteinD} yields the constraint equation
\begin{equation}
    \label{eq:constraint}
    e^{2A} \left[\frac{3 A'' + 6 (A')^2}{r_c^2} - \kappa^2 \mathcal{L}_\text{mat}\right] = \frac{\rho}{M_P^2} \,,
\end{equation}
implying that the left hand side of~\eqref{eq:constraint} is constant throughout the bulk. We note that the constraint does not involve integration over the internal space but has to hold pointwise. This reflects the fact that the energy density on each 4d slice is the same throughout the bulk, up to the factor $e^{2A}$.

We can now insert the constraint equation back into the action~\eqref{eq:Daction} in order to see that the result gives a consistent 4d theory. Doing so gives
\begin{equation}\begin{aligned}
    S_5 &=  \int d^5 x r_c \sqrt{- g} e^{4A}  \left[ \frac{1}{2 \kappa^2}  \left( e^{-2A} R_4  - \frac{(8 A'' + 20 (A')^2 )}{r_c^2} \right) + \mathcal{L}_\text{mat} \right] 
    \\ 
    &=  \int d^5 x r_c \sqrt{- g}\left[ \frac{e^{2A}}{2\kappa^2}\left( R_4  - \frac{2\rho}{M_P^2} \right) - \frac{e^{4A}}{r_c\kappa^2} \left( A'' + 4(A')^2 \right) \right] 
    \\ 
    &=  \int d^5 x r_c \sqrt{- g}\left[ \frac{e^{2A}}{2\kappa^2}\left( R_4  - \frac{2\rho}{M_P^2} \right) - \frac{1}{r_c\kappa^2} \partial_\phi (A'e^{4A})  \right] 
    \label{eq:dim_reduction}
\end{aligned}\end{equation}
Matching to the 4d theory requires that the 4d Planck scale, $M_P$, and $\kappa$ are related by the volume factor
\begin{equation}\label{eq:4daction}
    M^{2}_P = \frac{r_c}{2\kappa^2}  \int d\phi \, e^{2A} \, .
\end{equation}
After reinstating the boundary terms and combining those with the total derivatives in~\eqref{eq:dim_reduction}, the integration over $\phi$ leads to the 4d action:
\begin{align}
    S_4 = \int d^4x \sqrt{-g} \left( M_P^2 R_4 - 2\rho \right)
    - \sum_i \int_{B_i} d^4x \sqrt{-g_i} e^{4A} \left[ \pm \frac{6A'}{r_c \kappa^2} + \lambda_i(\Phi) \right] \,. 
\end{align}
As we will show in later sections, these boundary terms are proportional to the boundary conditions satisfied by the warp factor, and so vanish on-shell. We therefore see how requiring a static warped product geometry leads to an energy density which is the same on each bulk slice (again up to the factor $e^{2A}$). This guarantees a consistent four-dimensional cosmological term/Hubble constant at every point along the fifth dimension and a consistent effective theory. 
\\

Because it will be useful as a reference and nicely illustrates the above constraint, we review the solutions here for pure 5d AdS with only gravity in the bulk.  In section~\ref{sec:5dmodel} we solve the equations of motion for the full 5d model including a stabilizing field.  We consider a five-dimensional gravitational action with negative cosmological constant $\Lambda_5 = - 6 k^2$ coupled to two three-branes with tensions $\lambda_\alpha$. In the notation of our five-dimensional action \eqref{eq:Daction} this corresponds to 
\begin{equation}
    \mathcal{L}_\mathrm{mat} = - \frac{6 k^2}{\kappa^2} \,,
\end{equation}
without any additional five-dimensional bulk fields.
Also, the brane potentials $\lambda_i$ are constant as given below.

The four-dimensional metric is either Minkowski, de Sitter or Anti-de Sitter space-time, corresponding to an energy density
\begin{equation}
    \rho = 3 \Lambda M_P^2 \,,
\end{equation}
with $\Lambda$ the four-dimensional cosmological constant. Depending on the sign of $\Lambda$ the warp factor is given by~\cite{Karch:2000ct}
\begin{equation}
A(\phi) = - k \rc \left(c - \left|\phi\right|\right) \,,
\end{equation}
for the Minkowski case,
\begin{equation}
A(\phi) = \log \left[\frac{\sqrt{\Lambda}}{k} \sinh \bigl( k \rc (c - |\phi|)\bigr)\right] \,,
\end{equation}
for the dS case,
and
\begin{equation}
A(\phi) = \log \left[\frac{\sqrt{-\Lambda}}{k} \cosh \bigl( k \rc (c - |\phi|)\bigr)\right] \,,
\end{equation}
for the AdS case, respectively.

Consistently realizing these solutions when there are two branes \cite{DeWolfe:1999cp} requires appropriate boundary conditions, where Minkowski space requires a  vacuum critical value, and a nonzero c.c.  requires tensions detuned from that value,
\begin{equation} \label{eq:tuned_sols}\begin{aligned}
 \mathrm{Mink}_4 : \quad \Lambda = 0 & :
    \quad 
    \lambda_1 = - \lambda_2 = \frac{6 k}{\kappa^2} \,, \\
 \mathrm{dS}_4 : \quad \Lambda > 0 & :
    \quad 
    \lambda_1 = \frac{6k}{\kappa^2} \coth \bigl( k \rc c \bigr) \,, \quad
    \lambda_2 = -\frac{6k}{\kappa^2} \coth \bigl( k \rc (c - \pi) \bigr) \,, \\
 \mathrm{AdS}_4 : \quad \Lambda < 0 & :
    \quad
    \lambda_1 = \frac{6k}{\kappa^2} \tanh \bigl( k \rc c \bigr) \,, \quad
    \lambda_2 = -\frac{6k}{\kappa^2} \tanh \bigl( k \rc (c - \pi) \bigr)
\end{aligned}\end{equation}
Expanding for small $\Lambda$, we see that the boundary condition in the IR has two additional warp factor enhancements, since the $\Lambda$ piece of the warp factor is multiplied by a growing function of $\phi$.  We will see a similar enhancement for the change in IR boundary conditions when we consider UV perturbations later on.

We can easily check that all three cases satisfy the constraint equation \eqref{eq:constraint}.
Starting with the Minkowski case, we compute 
\begin{equation}
    \frac{3}{\rc^2} A''(r) + \frac{6}{\rc^2} A'(r)^2 - 6 k^2 = 0 = 3 \Lambda e^{-2A} \,,
\end{equation}
as well as for the de Sitter case
\begin{equation}
\frac{3}{\rc^2} A''(r) + \frac{6}{\rc^2} A'(r)^2 - 6 k^2 = 3 k^2 \sinh^{-2} \bigl( k \rc (c - |\phi|)\bigr) = 3 \Lambda e^{-2A} \,,
\end{equation}
and for the Anti-de Sitter case
\begin{equation}
\frac{3}{\rc^2} A''(r) + \frac{6}{\rc^2} A'(r)^2 - 6 k^2 = -3 k^2 \cosh^{-2} \bigl( k \rc (c - |\phi|)\bigr) = 3 \Lambda e^{-2A} \,.
\end{equation}
In the following we parametrize the metric so that there is a single four-dimensional cosmological constant throughout the bulk when we solve Einstein's equations.

\section{5d Model with a Stabilized Radion}

\label{sec:5dmodel}

In the previous section we assumed a static 5d theory, which requires tuned boundary tensions as in equation~\eqref{eq:tuned_sols}. At this level, any change in the brane tensions would destabilize the extra dimension and lead to time-dependent solutions, such as those considered in~\cite{Karch:2020iit}. We now add a stabilizing field~$\Phi$, which allows for stable solutions for a range of tensions. By solving the full system, we can determine how the metric adjusts when there is a localized perturbation to the energy, either in the UV or in the IR, and later demonstrate consistency with the EFT we will derive in later sections. This is of particular interest when we have IR supersymmetry breaking, for example, to see how a consistent solution develops throughout the bulk and in the UV. Here we perturb around a flat 4d metric but we will see the answer is readily generalizable. 

The original GW paper~\cite{Goldberger:1999uk} and most subsequent papers solved  the $\Phi$ equation of motion but neglected the backreaction on the gravitational metric. This is sufficient to determine $r_c$ in the 4d EFT. However, consistently accounting for boundary perturbations in the 5d picture and keeping track of how energy gets distributed requires including the back-reaction on the warp factor, even for small GW potential. In section~\ref{subsec:bulk} we derive the bulk field profiles for a general cosmological constant where we include the backreaction from $\Phi$. Then in section~\ref{subsec:boundaries} we will consider various choices of the GW potential on the boundaries and discuss which of them lead to stable solutions. We then consider two benchmark models with different choices of boundary potentials in sections~\ref{subsec:quadratic} and~\ref{subsec:linear}. In each case, we will give the leading order shift to the warp factor and GW field from a perturbation in either the IR or UV brane tension. From this we can determine the shift in value of the radion and the 4d cosmological constant. We will use these results in section~\ref{sec:effectivepotential}  to derive the full 4d effective potential for our models.

\subsection{Bulk Profiles}

\label{subsec:bulk}

As in the previous section, we consider a five-dimensional gravitational theory with co-dimension one branes $B_i$, coupled to a scalar field $\Phi$. The action is given by
\begin{equation} 
    S = \int d^5x \sqrt{-g^{(5)}}  \left[\frac{1}{2 \kappa^2} R - \frac12  (\partial \Phi)^2 - V(\Phi)  \right] - \sum_i \int_{B_i} d^4x \sqrt{-g_i} \lambda_i(\Phi) \, .
    \label{eq:5daction}
\end{equation}
For a metric of the form~\eqref{eq:metricansatz}, where 4d slices are described by a maximally symmetric spacetime with c.c. $\Lambda$ (i.e.\ $R_{\mu\nu} = 3 \Lambda g_{\mu\nu}$) the bulk equations of motion read~\cite{DeWolfe:1999cp}
\begin{equation}\begin{aligned}\label{eq:bulkeoms}
    \Phi'' + 4 A' \Phi' - \rc^2 \frac{\partial V(\Phi)}{\partial \Phi} &= 0 \,, \\
    A'' + \rc^2 \Lambda e^{-2A} + \frac{\kappa^2}{3} \Phi'^2 &= 0 \,, \\
    A'^2 - \rc^2 \Lambda e^{-2A}- \frac{\kappa^2}{12} \Phi'^2 + \frac{\rc^2 \kappa^2}{6} V(\Phi) &= 0 \, .
\end{aligned}\end{equation}
In equation~\eqref{eq:bulkeoms} we have assumed that $\Phi$ is independent of the four-dimensional co-ordinates, and denote $\phi$ derivatives by a prime.

We will consider a simple bulk potential consisting of a five-dimensional, negative cosmological constant and a mass-term for $\Phi$,\footnote{Keeping just a mass term for $\Phi$ allows us to write analytic solutions for the background profile for $\Phi$, which is why we make that assumption here. Higher order terms in the bulk potential can also be considered, as was done in ref.'s~\cite{Lewandowski:2001qp, Chacko:2013dra, Chacko:2014pqa, Mishra:2024ehr}.}
\begin{equation}
    V = - \frac{6 k^2}{\kappa^2} + 2 m^2 \Phi^2 \,,
    \label{eq:Vbulk}
\end{equation}
together with two branes at $\phi_1=0$ and $\phi_2=\pi$ with brane-localized potentials
\begin{align}
    &\lambda_1 = \frac{6 k}{\kappa^2} + \Delta T_{\uv} +\lambda_{\uv} (\Phi) \,, 
    &&\lambda_2 = - \frac{6 k}{\kappa^2} + \Delta T_{\ir} + \lambda_{\ir} (\Phi) \,.
    \label{eq:branepotentials}
\end{align}
The bulk equations of motion have to be amended by the boundary or jump conditions at the locations $\phi_i$ of the branes,
\begin{align}\label{eq:boundaryconditions}
    & \left. A' \right |^{\phi_i + \epsilon}_{\phi_i - \epsilon} = - \frac {\kappa^2}{6} \rc \lambda_i \bigl(\Phi(\phi_i)\bigr) \,, 
    && \left. \Phi' \right |^{\phi_i + \epsilon}_{\phi_i - \epsilon} = \rc \frac{\partial\lambda_i}{\partial \Phi} \bigl(\Phi(\phi_i)\bigr) \, ,
\end{align}
where $i = 1, 2$ labels the two branes.

In general, we will not be able to solve the system~\eqref{eq:bulkeoms} analytically. However, assuming that the backreaction induced by a non-vanishing $\Phi$ and/or $\Lambda$ is sufficiently small, the backreaction can be treated in a first-order approximation. This amounts to assuming that $\Phi \ll M_5^{3/2} $ throughout the bulk and that $|\Lambda| \ll k^2 e^{-2kr_c \pi}$. The zeroth-order solution that we are expanding around then corresponds to 
\begin{equation}
    A(\phi) = - k \rc \phi \, ,
\end{equation}
with $\Phi = \Lambda =0$. Plugging this into the first equation in~\eqref{eq:bulkeoms} then gives the general solution for $\Phi$
\begin{equation}\label{eq:Phisol}
    \Phi(\phi) =  C_1 \, e^{2k \rc \phi (1+ \sqrt{1+\epsilon})} + C_2 \, e^{2k \rc \phi (1- \sqrt{1+\epsilon})} \,,
\end{equation}
with integration constants $C_{1,2}$ to be determined by the boundary conditions~\eqref{eq:boundaryconditions}. Here we have introduced the parameter $\epsilon = m^2/k^2$.\footnote{We note that this convention differs by a factor of $4$ from some other works, see e.g.~\cite{Creminelli:2001th}.} We will take $\epsilon$ to be a small parameter in order to naturally generate a hierarchy between the UV and IR scales.

We next determine the backreaction on the profile of the warp factor $A(\phi)$ by expanding it as
\begin{equation}
   A(\phi) = - k \rc \phi + \delta A(\phi) \,.
\end{equation}
To capture the leading order backreaction effects we can expand \eqref{eq:bulkeoms} to linear order in $\delta A(\phi)$ and $\Lambda$ and to quadratic order in $\Phi$. The resulting equation for $\delta A(\phi)$ reads
\begin{equation}
    \delta A' (\phi) = - e^{2k \rc \phi} \frac{ \rc }{2 k }  \Lambda - \frac{\kappa^2}{24 k \rc}  \Phi'(\phi)^2 + \frac{\epsilon k \rc \kappa^2}{6} \Phi(\phi)^2 \,.
\end{equation}
After inserting~\eqref{eq:Phisol} this can be straightforwardly integrated and we find
\begin{align}
    A(\phi) = - k \rc \phi - \frac{\Lambda}{4 k^2} e^{2 k \rc \phi}
    - \frac{\kappa^2}{12} e^{4 k \rc \phi} \left(C_1^2 e^{4 k \rc \phi \sqrt{1+\epsilon}} + C_2^2 e^{-4 k \rc \phi \sqrt{1+\epsilon}} - 2 C_1 C_2 \epsilon \right) + C_3 \,, 
\end{align}
keeping only the leading-$\epsilon$ piece for each of the terms. The integration constant $C_3$ does not have any physical meaning and can be absorbed into a rescaling of $\Lambda$ or the location of the UV brane~\cite{DeWolfe:1999cp}. Here, we chose to fix it by setting $A(0) = 0$,
\begin{equation}
    C_3 = \frac{\Lambda}{4k^2} + \frac{\kappa^2}{12} \left(C_1^2 + C_2^2 - \epsilon C_1 C_2 \right) \,.
\end{equation}

Finally, as it will become useful in subsequent sections, we also show how to express $C_1$ and $C_2$ in terms of the values of $\Phi$ at the positions of the branes,
\begin{equation}
    C_1 = \frac{\Phi(0)- \Phi(\pi) e^{-2k \pi \rc (1- \sqrt{1+\epsilon})}}{1 - e^{4k \pi \rc \sqrt{1+\epsilon}}} \,, \qquad
    C_2 = \frac{\Phi(0)- \Phi(\pi) e^{-2k \pi \rc (1+ \sqrt{1+\epsilon})}}{1 - e^{-4k \pi \rc \sqrt{1+\epsilon}}} \,.
\end{equation}
The values of $\Phi(0)$ and $\Phi(\pi)$ depend on the boundary potentials for $\Phi$, $\lambda_\uv (\Phi)$ and $\lambda_\ir (\Phi)$ through the boundary conditions~\eqref{eq:boundaryconditions}. In the following subsection we find $\Phi(0)$ and $\Phi(\pi)$ for some well-motivated examples of boundary potentials.

\subsection{Boundary conditions}

\label{subsec:boundaries}

The full solution for the coupled field and geometry is characterized by four constants of integration, $\Phi(0)$, $\Phi(\pi)$, $\rc$, and $\Lambda$, which are fixed by the four boundary conditions~\eqref{eq:boundaryconditions} for $A'$ and $\Phi'$ on the two branes. Determining these constants requires a choice of boundary potential for the GW field on each brane. In the following we will consider various combinations of different brane potentials:
\begin{equation}
\begin{aligned}
    \text{quartic:}
    &\qquad \lambda_{i}(\Phi) = \frac{\alpha_{i}}{2k^2} \left(\Phi^2 - v_i^2 \right)^2 \,, \\
    \text{quadratic:}
    &\qquad   \lambda_{i}(\Phi) = \alpha_i k \left(\Phi - v_i \right)^2 \,, \\
    \text{linear:}
    &\qquad \lambda_i(\Phi) = \alpha_i k^{5/2} \Phi \, , \\
\end{aligned}
    \label{eq:boundarypotentials}
\end{equation}
where $i = \uv, \ir$ and the factors of $k$ are chosen such that the couplings $\alpha$ are dimensionless. We also consider different signs of $\epsilon$, where $\epsilon$ positive/negative corresponds to $\Phi$ which decreases/increases in the IR.

Not all possible boundary potentials lead to a consistently stabilized radion, and which ones do depends on the sign of $\epsilon$. Solving the boundary conditions for $A'$ and $\Phi$ determines the critical points of the potential, which could be either minima or maxima. Determining if a solution is a stable minimum therefore requires constructing the effective potential for the radion or equivalently considering fluctuations around the background solution. The couplings in the potential~\eqref{eq:Veff_general} are calculated in section~\ref{sec:effectivepotential}, but we summarize the relevant results here.

Defining the radion as $\varphi = e^{-k \pi r_c}$, the effective potential for $\varphi$ has the general form
\begin{align}
    V_{\rm eff} (\varphi) = \lambda_4 \varphi^4 + \lambda_{4+\epsilon} \varphi^{4+\epsilon} + \lambda_{4+2\epsilon} \varphi^{4+2\epsilon} + \mathcal{O}(\varphi^8)\, ,
    \label{eq:Veff_general}
\end{align}
where the couplings depend on $\epsilon$ and the parameters of the boundary potentials. $\lambda_4$ can be nonzero owing to the GW  potential and because of a non-vanishing detuning $\delta T_\ir$ of the IR brane tension. For either sign of $\epsilon$, a stable minimum is possible only if 
\begin{align}
    \lambda_4  \text{ and } \lambda_{4+2\epsilon} >0 , \ \ \lambda_{4+\epsilon}<0 \,.
\end{align}
Another possibility is if $\lambda_{4+\epsilon} >0$ and the leading term for small $\varphi$ is negative (this is the $\lambda_4$ term if $\epsilon>0$, or the $\lambda_{4+2\epsilon}$ term if $\epsilon<0$). The potential in eq.~\eqref{eq:Veff_general} ignores higher order terms which start at $\mathcal{O}(\varphi^8)$, which allows for the possibility that $V_{\rm eff}$ (in our approximation scheme) is unbounded but is stabilized by higher-order terms. As we want to focus on models that are stabilized with a large hierarchy, $\varphi \ll 1$, we require the potential~\eqref{eq:Veff_general} be positive at large $\varphi$. With these requirements, we state which models can and cannot give a stable minimum:
\begin{itemize}
    \item If there are linear boundary potentials in both the UV and IR there is no non-zero minimum for $\varphi$. In this case both $\lambda_{4+\epsilon}, \, \lambda_{4+2\epsilon}$ are proportional to $ - 1/\epsilon$, while $\lambda_4$ can have either sign depending on $\delta T_\ir$. Therefore the only possibilities are that there is a maximum, or all of the $\lambda$'s have the same sign, leading to a runaway.
    
    \item When there are either quadratic or quartic potentials on both branes, $\lambda_{4+\epsilon} < 0$ and $\lambda_{4+2\epsilon} >0$. When $\epsilon >0 $ the quartic is also positive for $\delta T_\ir =0$, so a minimum can be found for a critical IR brane tension. This is the case we consider in section~\ref{subsec:quadratic}. In the $\epsilon<0$ case $\lambda_4$ is negative when $\delta T_\ir =0$, which would lead to a runaway as $\varphi \to 1$. By adding a nonzero $\delta T_\ir$, $\lambda_4$ can be made positive, leading to a stable minimum.
    
    \item Taking a quadratic or quartic boundary potential in the UV, but a linear boundary potential in the IR, we find that $\lambda_{4+2\epsilon} \propto - \epsilon$. This means that the $\epsilon>0$ case does not work unless the potential can be stabilized by higher order terms in the potential. For negative $\epsilon$ we find that $\lambda_{4+\epsilon}\propto \alpha_\ir$, so is negative if $\aIR<0$. To have a stable solution for large $\varphi$ (again in the absence of higher order terms in $V_{\rm eff}$) we also require a nonzero $\delta T_\ir$ so that $\lambda_4$ is positive. This is the case we consider in~\ref{subsec:linear}.

    \item Finally, we can also consider a linear boundary potential in the UV, and a quadratic or quartic boundary potential in the IR. In both cases $\lambda_{4+\epsilon} \propto \aUV/\epsilon$ and $\lambda_{4+2\epsilon}, \lambda_{4}$ are positive. For positive $\epsilon$ this means we should take $\aUV<0$ to generate a minimum, while for negative $\epsilon$ a minimum requires $\aUV>0$. We do not consider these possibilities in this work.
\end{itemize}

\subsubsection{Relation to the Goldberger-Wise Model}

The original GW model~\cite{Goldberger:1999uk} involved a quartic potential on both branes where the limit $\alpha_{\uv / \ir} \to \infty$ was taken to set $\Phi(0) \to \vUV \, , \Phi(\pi) \to \vIR$. This is insufficient for out aims, as we need to include $1/\alpha$ corrections to study how the solutions respond to changes in parameters, even though the shifts in $\Phi(0), \Phi(\pi)$ are suppressed by $1/\alpha$ and therefore vanish in the GW limit.  Looking at the boundary condition in the UV for $\Phi' (0)$ in the GW model:\footnote{The boundary condition in the IR follows a similar story.}
\begin{align}
    \Phi'(0) &= \frac{2 r_c \aUV}{k^2} \Phi (0) \left(\Phi(0)^2 - v_\uv^2 \right) \, ,
\end{align}
it seems that taking $\aUV \to \infty$ and $\Phi(0) \to \vUV$ sets $\Phi'(0) = 0$. However, if we include subleading corrections to $\Phi(0)$,
\begin{align}
    \Phi(0) = \vUV + \frac{\delta \Phi}{\aUV} + \ldots \, ,
\end{align}
where $\delta \Phi$ is independent of $\aUV$, we find that 
\begin{align}
    \Phi'(0) &\to \frac{2 r_c \vUV^2}{k^2} \delta \Phi \, .
\end{align}
This means that $\Phi'$ can be nonzero on the boundaries even in the $\alpha \to \infty$ limit, but this is clear only when taking the large $\alpha$ limit at the end of the calculation. In any case we will consider arbitrary $\alpha$ values on both branes, taking the large $\alpha$ limit in some examples only to simplify expressions. At finite $\alpha$, the quartic potential becomes difficult to work with as the boundary conditions are cubic in $\Phi$. In order to obtain analytic results without making unnecessary approximations, we focus on the quadratic and linear potentials in the rest of this section. In any case, for small perturbations around the GW solution, the quadratic potential gives a reliable approximation to the quartic. For large but finite  $\alpha$ we expect $\Phi(0) \approx \vUV \, , \Phi(\pi)\approx \vIR$, so we can expand the quartic brane potentials in equation~\eqref{eq:boundarypotentials} around these values as follows:
\begin{equation}\begin{aligned}
    \lambda_{\uv}(\Phi) &= \frac{2}{k^2} \aUV \vUV^2 (\Phi - \vUV)^2 + \mathcal{O}\left[(\Phi - \vUV)^3 \right] \,, \\
     \lambda_{\ir}(\Phi) &= \frac{2}{k^2} \aIR \vIR^2 (\Phi - \vIR)^2 + \mathcal{O}\left[(\Phi - \vIR)^3 \right] \, .
\end{aligned}\end{equation}
Therefore we can approximate the GW model by the quadratic/quadratic brane potentials if we make the replacement
\begin{equation}
        \alpha_{\uv / \ir} \rightarrow \frac{2}{k^3} \, \alpha_{\uv / \ir} \, v_{\uv / \ir}^2 \,.
\end{equation}
\\

In the following sections we construct the leading order solutions and then show how the GW field (and the metric) adjust in the presence of a perturbative change in tension on either of the two branes. This could occur dynamically in the early universe, for example, due to a phase transition on either of the branes or could, as alluded to in the introduction, originate from supersymmetry-breaking energy. 

In section~\ref{subsec:quadratic} we consider quadratic brane potentials. In section~\ref{subsec:linear} we consider a linear potential on the IR brane with a quadratic UV brane potential, as is often discussed in a holography-motivated context.

\subsection{Solution with Quadratic/Quadratic Boundary Conditions}
\label{subsec:quadratic}

As our first benchmark case we consider $\epsilon>0$ and quadratic potentials on the two branes:
\begin{align}
    &\lambda_{\uv}(\Phi) = \aUV k \left(\Phi - \vUV \right)^2 \,,
    &&\lambda_\ir(\Phi) =  \aIR k \left(\Phi - \vIR \right)^2 \,,
\end{align}
With the solution~\eqref{eq:Phisol} we can solve the resulting boundary conditions \eqref{eq:boundaryconditions} for $\Phi'$ at $\phi = 0$ and $\phi = \pi$ in terms of $\Phi(0)$ and $\Phi(\pi)$.
At leading order in $e^{-k \pi r_c}$  we find\footnote{We note that although we can drop higher order terms in $e^{-k \pi r_c}$, terms of order $e^{- \epsilon k \pi r_c}$ can be $\mathcal{O}(1)$ and can't be ignored.}
\begin{equation}\begin{aligned}
\label{eq:GWboundaries}
    \Phi(0) &= \frac{ \aUV \vUV}{\aUV + \epsb } \, , \\
    \Phi(\pi) &= \frac{1}{\aIR + 4 + \epsb} \left( \aIR \vIR 
    + (4+2\epsb) \frac{ \aUV  \vUV e^{- \epsb  k \pi \rc}}{ \aUV + \epsb } \right) \, ,
\end{aligned}\end{equation}
where to simplify the notation we have defined 
\begin{align}
    \epsb = 2( \sqrt{1+\epsilon} - 1) \, ,
    \label{eq:epsb_def}
\end{align}
noting that $ \bar \epsilon \simeq \epsilon $ for small $\epsilon$. As expected, in the limit $\left|\aUV\right| \rightarrow \infty$, $\left|\aIR\right| \rightarrow \infty$ these expressions reduce to $\Phi(0) = \vUV$ and $\Phi(\pi) = \vIR$.  However, the suppressed terms are essential for solving the scalar boundary conditions.

To fully specify the theory, we need the boundary tensions. In the presence of the stabilizing potential, there are generically GW-potential-dependent contributions to the cosmological constant that we tune to zero in the initial unperturbed solution. This means that before any additional perturbation,  we have an initial state in which parameters are tuned to yield the four-dimensional c.c. $\Lambda = 0$. A zero c.c.\ then requires  detuning the brane tension away from the (vacuum) critical value so that the additional brane tension cancels such extra contributions. Here we choose to detune the tension of the UV-brane by allowing for a non-zero $\Delta T_\uv$, while setting $\Delta T_\ir=0$. 

Solving for the boundary condition for $A'(0)$ in equation~\eqref{eq:boundaryconditions} with $\Lambda = 0$ gives the critical detuning $\Delta T_{\uv}^0$:
\begin{equation}
    \Delta T_{\uv}^0  =  -  \frac{ \aUV k \vUV^2 \epsb }{\aUV + \epsb} 
    + \Delta T_{\uv}^{(4)} \, .
    \label{eq:UVcrit_tuning}
\end{equation}
where $\Delta T_{\uv}^{(4)}$ is a subleading correction suppressed by $e^{-4k\pi r_c}$.
In the large $\aUV$ limit the leading term reduces to
\begin{equation}
    \Delta T_{\uv}^0  = - k \vUV^2 \eps \,.
\end{equation}
It is important to keep the higher order terms (proportional to $e^{-k \pi r_c}$) as they are suppressed by the same number of warp factors as a c.c. generated from a detuned brane tension in the IR. This contribution is denoted by $\Delta T_{\uv}^{(4)} $ in equation~\eqref{eq:UVcrit_tuning} and is given by
\begin{equation}\label{eq:UVcrit_tuning4}
\begin{aligned}
    \Delta T_{\uv}^{(4)}  &=  e^{-4 k \pi r_c} \frac{\aIR k \vIR^2 \epsb \left(\sqrt{\epsb \aIR (\aIR+4) (4+\epsb)}+\epsb(4+ \epsb) \right)}{(2 + \epsb) (\aIR-\epsb ) (\aIR+4 + \epsb)}
    \, ,
    \\
    &= e^{-4 k \pi r_c}  \left( k \vIR^2 \epsilon^{3/2} \sqrt{ \frac{\aIR}{\aIR+4}}  + \mathcal{O}(\epsb^2) \right)\, .
\end{aligned}    
\end{equation}
We note that this subleading term in the critical UV brane tension corresponds to the cosmological constant found in ref.~\cite{Bellazzini:2013fga}. Due to the stabilization mechanism the 4d cosmological constant takes the same value across the entire extra dimension so it is sensitive to both UV and IR contributions. We choose to tune the UV contribution so that the entire system is described by 4d Minkowski space before perturbing the UV and IR brane tensions and determining the cosmological constant that would subsequently be generated. Because ref.~\cite{Bellazzini:2013fga} neglected the leading IR contribution to the c.c.\ when solving for the UV detuning, they were left with an apparent nonzero c.c.\ suppressed by $\epsilon$ and four warp factors. This is simply because the tuning in the presence of the stabilizing potential did not account for this suppressed energy contribution.

Finally, we can determine $r_c$ by solving the $A'(\pi)$ boundary condition to leading order in $e^{-k \pi r_c}$:
\begin{equation}
\begin{aligned}
\label{eq:rcsolution}
    e^{ - \epsb k \pi \rc} &= \frac{\vIR}{\vUV}
    \left[\frac{ (\aUV + \epsb) (\aIR(4+\epsb) \pm \sqrt{\epsb (4+\epsb) \aIR (\aIR + 4)} )}{ (4+2\epsb)\aUV \left( \aIR - \epsb \right)}
    \right] \\
    &\simeq \frac{\vIR}{\vUV} \left[ 1 \pm  \frac12 \sqrt{\frac{(\aIR+4)\epsb}{\aIR}} + \frac{( 4(\aIR +\aUV) - \aIR \aUV) \epsb}{4 \aIR \aUV} + \mathcal{O}(\epsb^{3/2})\right]
    \,.
\end{aligned}
\end{equation}
The two solutions in \eqref{eq:rcsolution} are both critical points of the potential. However, while the $+$ solution is a local minimum, the $-$ solution corresponds to a local maximum so we neglect it.

At this stage we have found a stable solution that leads to a Minkowski metric in 4d in the presence of the stabilizing potential. Our goal now is to study the response of the system to a small shift in the tension of either or both of the two branes. In this way we can determine how the metric and GW field react in the presence of a perturbation to generate a uniform 4d c.c.~throughout the bulk and ultimately derive the correct low-energy effective theory. So we set
\begin{align}
    &\Delta T_{\uv} = \Delta T_{\uv}^0 + \delta T_{\uv}  \,, 
    &&\Delta T_{\ir} = \delta T_{\ir} \,,
\end{align}
and compute the first-order perturbations around the previously determined solution. We solve for the deformations caused by $\delta T_\uv, \, \delta T_\ir$ separately, but when they are both present the perturbations all add so long as we remain in the perturbative regime.

We first perturb the tension of the UV-brane, setting $\delta T_\ir = 0$. In this case we find the cosmological constant at leading order in $e^{-k \pi \rc}$ by solving the boundary condition for $A'(0)$
\begin{equation}\label{eq:deltaLambdaUVquad}
    \Lambda = \frac{k \kappa^2}{3} \delta T_{\uv} \,.
\end{equation}
The shift in the critical radius, $r_c$, to leading order in $\epsilon$ is
\begin{align}
    \delta r_c \simeq - \frac{\delta T_\uv e^{2k\pi r_c}}{6 \pi k^2 \vIR^2 \epsilon^{3/2}} \sqrt{\frac{4+\aIR}{\aIR}} \, ,
    \label{eq:deltar_UV}
\end{align}
while the boundary values of the GW field shift as
\begin{align}
    & \delta \Phi(\pi) \simeq
    \frac{2 \delta T_\uv e^{2k\pi r_c}}{k\vIR \sqrt{\aIR\epsilon  (4+\aIR) }} \, ,
    && \delta \Phi(0) \simeq 
    -\frac{2\delta T_\uv e^{-2k\pi r_c}}{3 k \epsilon \aUV \vUV } \sqrt{\frac{\aIR}{4+\aIR}}\, .
    \label{eq:deltaGW_UV}
\end{align}
The condition for us to stay within the regime of small backreaction is 
\begin{align}
   \delta \Lambda \ll k^2 e^{-2 \pi k r_c} 
   \qquad
    \implies
     \qquad
    \delta T_{\uv} \ll \frac{3k}{\kappa^2} e^{-2 \pi k r_c}\, ,
    \label{eq:TUVbound}
\end{align}
which comes from inspecting the third equation of motion in~\eqref{eq:bulkeoms} in the IR where the backreaction from $\Lambda$ is largest. This implies that the $\delta T_\uv $ appearing in equations \eqref{eq:deltar_UV}~\&~\eqref{eq:deltaGW_UV} is at most $\mathcal{O} \left(e^{-2k\pi r_c}\right)$, so $\delta r_c$ and $\delta \Phi$, though enhanced by powers of $e^{k\pi r_c}$, are not outside the perturbative regime. Notice that the shift in the radion (and the shift in GW field more generally) depends on the parameters of the stabilizing potential, but the shift in the cosmological constant does not.

The shift in $\Phi$ also leads to changes in the brane energy densities:
\begin{equation}
\label{eq:GWpotentialsUV}
\begin{aligned}
    \delta \lambda_\uv (\Phi) &= 
    \lambda_\uv (\Phi(0) + \delta \Phi(0)) - \lambda_\uv (\Phi(0))
    = \frac{4 \delta T_\uv e^{-2\pi kr_c} }{3(\aUV +\epsb)}  \sqrt{\frac{\aIR \epsb}{4+\aIR}}\, ,
    \\
    \delta \lambda_\ir (\Phi) &= 
    \lambda_\ir (\Phi(\pi) + \delta \Phi(\pi)) - \lambda_\ir (\Phi(\pi))
    \simeq \frac{8 \delta T_\uv e^{-2\pi kr_c}}{3(4+\aIR)} \, .
\end{aligned}
\end{equation}
Here we see that both $\delta \lambda_\uv$ and $\delta \lambda_\ir$ vanish in the limit of large $\aUV\, , \aIR$. This might seem inconsistent with the modified c.c., but is not because the boundary condition involves both the derivative of the warp factor and the brane tension. The warp factor adjusts due to the shift in the GW scalar to maintain consistency with the change in the 4d c.c. Although the energy density on the IR brane is unchanged in this limit, the warp factor is not. The change in $A'(\theta)$ on each of the branes allows the boundary conditions to be satisfied:
\begin{equation}
\label{eq:GWpotentialsUVB}
\begin{aligned}
    \delta A'(0) &\simeq - \frac{4 \kappa ^2 \rc \delta T_\uv e^{-2\pi kr_c} }{9 (\aUV+\epsilon )} \, ,
    \\
    \delta A'(\pi) &\simeq -\frac{\kappa ^2 \aIR \delta T_\uv e^{2\pi kr_c}}{9(4+\aIR)} 
    \, .
\end{aligned}
\end{equation}

We now turn to the case where we perturb only the tension of the IR-brane, i.e. $\delta T_{\uv} = 0$. For non-vanishing $\delta T_{\ir}$ the critical radius is given by
\begin{equation}\begin{aligned}
    e^{ - \epsb k \pi \rc} &= 
    \frac{(\aUV+\epsb ) \left[\aIR k \vIR (4 + \epsb) 
    + \sqrt{4k\delta T_{\ir} \left( \epsb (4 + \epsb) - \aIR( 4 +\aIR) \right) 
    +\aIR k^2 \vIR^2 \epsb (4+\epsb) \left( 4+ \aIR \right)}\right]}{2 k \aUV  \vUV(2 + \epsb) (\aIR-\epsb)} \\
    &\simeq \frac{\vIR}{\vUV }+\frac{\sqrt{k \vIR^2 \epsb -  \delta T_{\ir} }}{2k^{1/2} \vUV} \, ,
    \label{eq:minimum1}
\end{aligned}\end{equation}
where in the final step we have taken the large $\alpha$ limit to simplify the expression. For a solution to exist the square root must be real, which implies that\footnote{This is what caused the apparent runaway that was discussed in refs.~\cite{Bena:2018fqc, Randall:2019ent, Dudas:2019pls, Lust:2022xoq}.}
\begin{equation}
    \delta T_{\ir} \leq \frac{k \vIR^2 \aIR(4+\aIR) \epsb  (4+\epsb)}
    {4\left(\aIR(4+\aIR)- \epsb(4+\epsb)\right)} \simeq k \vIR^2  \epsb  \, .
    \label{eq:TIRbound1}
\end{equation}
There is also a lower bound on $\delta T_\ir$ that comes from requiring  $r_c>0$. To leading order in $\epsb$ this is
\begin{align}
    \delta T_\ir \gtrsim -\frac{4 \aIR k (\vIR - \vUV)^2}{\aIR+4}  \, .
    \label{eq:TIRbound2}
\end{align}
We note that this bound should not be taken as precise as we expect the expansion in $e^{-k \pi r_c}$ to break down before $r_c = 0$. However we include it to highlight that $\delta T_\ir$ cannot be made arbitrarily negative.

For a sufficiently small change in tension the critical radius shifts by
\begin{align}
    \delta r_c = \frac{\delta T_\ir}{4k^2 \pi \vIR^2 \epsilon^{3/2}} \sqrt{\frac{\aIR + 4}{\aIR}} \, ,
    \label{eq:deltarcquad}
\end{align}
to leading order in $\epsilon$. We note that due to the $\epsilon^{-3/2}$ enhancement it is possible to get large shifts in $r_c$ while satisfying the bounds~\eqref{eq:TIRbound1} and \eqref{eq:TIRbound2}. This corresponds to the fact that the radion mass is proportional to $\epsilon^{3/2}$, which can in part be understood as the radion is the pseudo-goldstone boson of scale invariance, where $\epsilon$ sets the strength of the breaking~\cite{Rattazzi:2000hs, Arkani-Hamed:2000ijo}. Solving the boundary condition for $A'(0)$, at leading order in $e^{-k \pi \rc}$ we find the leading order shift in the cosmological constant
\begin{equation}\label{eq:DLambdaIR}
    \delta \Lambda = \frac{k \kappa^2}{3} e^{-4k \pi \rc}  \delta T_{\ir} \, ,
\end{equation}
so all the dependence on the parameters of the brane potentials has dropped out and the c.c.\ is proportional to the warped energy density on the IR brane. We note that the next-to-leading term, $\delta \Lambda^{(2)}$, in both $\delta T_\ir$ and $\epsb$ scales as:
\begin{equation}\label{eq:deltaLambdaIRquad}
    \delta \Lambda^{(2)} \simeq - \frac{\kappa^2 (\delta T_{\ir})^2 e^{-4k \pi \rc}}{6 v_\ir^2 \epsb^{3/2} } \sqrt{ \frac{4 + \aIR}{\aIR}} \, ,
\end{equation}
so the expansion in $\delta T_\ir$ begins to break down at the point where equation~\eqref{eq:TIRbound1} is violated and the  solution is destabilized.

The result~\eqref{eq:DLambdaIR} is important and differs from the result for scalar gravity models~\cite{Coradeschi:2013gda, Agrawal:2016ubh}, which do not include a UV brane. These models find a vacuum energy that is suppressed by the small parameter $\epsilon$ relative to the expected value found in equation~\eqref{eq:DLambdaIR}. However, in models with a UV brane both the contributions to the c.c.\ from the UV and IR need to be considered together. In these cases it is typically assumed that the UV brane tension is tuned to cancel the UV-parameter-dependent contributions to the cosmological constant. We have seen that unless this cancellation includes the subleading $\Delta T_\uv^{(4)}$ term in equation~\eqref{eq:UVcrit_tuning4}, that a c.c. of order $ \sim \epsilon e^{-4k\pi r_c}$ is left uncancelled. This is the contribution from the IR GW field, and is the c.c.\ that was left uncancelled in~\cite{Bellazzini:2013fga}, and is the same term in the scalar gravity models that leads to a c.c.\ suppressed by $\epsilon$. In our case, we consider both the UV and IR tensions together when tuning the c.c.\ to zero initially before asking how things change when we perturb away from that solution.  

The boundary conditions for the GW field (equation~\eqref{eq:boundaryconditions}) don't depend directly on the brane tensions as they are set by derivatives of the boundary potentials. Therefore the expressions for $\Phi(0)$ and $\Phi(\pi)$ in equation~\eqref{eq:GWboundaries} remain the same after the perturbation. The UV boundary value therefore remains fixed to leading order in $e^{-4 k \pi r_c}$, with the leading correction due to the change in $r_c$:
\begin{align}
    \delta \Phi(0) =
    \frac{2 \delta T_{\ir}e^{-4 k \pi r_c}}
    {k \vUV \aUV (2 + \epsb)}
    \left( 1 + 2\sqrt{ \frac{\aIR}{\epsb (4+\epsb)(\aIR+4)} }\right) \,.
    \label{eq:delta_phiuv}
\end{align}
The shift in the IR value of the GW field is (to leading order in $\delta T_{\ir}$):
\begin{align}
   \delta \Phi(\pi) = 
   -\frac{2\delta T_{\ir}}{k \vIR \sqrt{\aIR (\aIR+4) \epsb (4+\epsb)}} \, .
\end{align}

In a holographic interpretation, the GW field, $\Phi$, corresponds to a coupling in the CFT that runs, breaking scale invariance~\cite{Rattazzi:2000hs, Arkani-Hamed:2000ijo}. The perturbation~\eqref{eq:delta_phiuv} we find on the gravity side then corresponds to the necessary change in the UV boundary condition so that the RG flow results in an IR theory with an extra energy density given by $\sim \delta T_\ir e^{-4k\pi r_c}$. In the holographic language, the constraint of consistent slicing simply means that the c.c. is present at all energy scales.

The leading shifts in the derivatives of the GW field are:
\begin{align}
   \delta \Phi'(0) &= \frac{r_c}{\vUV} \left(\sqrt{\frac{\aIR }{\epsb(4+\aIR)}} +1 + \mathcal{O(\epsb)} \right)  \delta T_{\ir}e^{-4 k \pi r_c} \,,
    \\
    \delta \Phi'(\pi) &= 
   -\frac{2r_c \delta T_{\ir}}{\vIR }
  \sqrt{ \frac{\aIR}{\epsb (4+\epsb)(\aIR+4) }}\, .
\end{align}
The corrections to the energy densities on each brane from the GW potentials, as defined in equation~\eqref{eq:GWpotentialsUV}, are:
\begin{equation}
\begin{aligned}
    \delta \lambda_\uv (\Phi) &= - 
    \frac{2 \delta T_\ir e^{-4k \pi r_c} \left((\aIR+4) \epsilon +\sqrt{\aIR (\aIR+4) \epsilon }\right)}{(\aIR+4) \aUV} \, ,
    \\
    \delta \lambda_\ir (\Phi) &= -
    \frac{4\delta T_\ir \left(\aIR (\aIR+4)- 2 \sqrt{\aIR (\aIR+4) \epsilon } \right)}{\aIR (\aIR+4)^2}\, ,
\end{aligned}
\end{equation}
where again both $\delta \lambda_\uv$ and $\delta \lambda_\ir$ vanish in the limit of large $\aUV\, , \aIR$.

\subsection{Quadratic UV Potential/ Linear IR potential}

\label{subsec:linear}

Motivated by holographic theories, we next consider a model with a linear boundary potential in the IR. Again we define our potential so that $\alpha$ is dimensionless:
\begin{align}
    \lambda_\ir = \aIR k^{5/2} \Phi \, .
\end{align}
As discussed in section~\ref{subsec:boundaries}, for this to lead to a stabilized radion requires $\epsilon < 0$ and $\aIR < 0$. This means that $\Phi$ grows in the IR, so this model has a natural interpretation as a technicolor-like theory with a scale generated due  to a coupling growing large in the IR. This model is nice in that it establishes a source term for the GW field in the UV but not in the IR so can more closely resemble a fundamental holographic theory.

However, this growing field alone is not sufficient to have a stable IR brane. Unlike the $\epsilon>0$ case considered above, there is a minimum IR detuning required for the radion potential to not have a runaway as $\varphi \to 1$. The quartic coupling in the effective potential receives contributions both from detuning in the IR and from the GW potential. Therefore the requirement is that the quartic coupling in the effective potential \eqref{eq:Veff_general}, $\lambda_4$, is positive:
\begin{align}
    \lambda_4 = \Delta T_\ir -  \frac{k^4 \aIR^2}{4(4+\epsb)} >0 \, ,
    \label{eq:mindetuning}
\end{align}
where we again define $\epsb = 2(\sqrt{1+\epsilon} - 1)$ as in the $\epsilon>0$ case. We will also find an upper bound on the tension:
\begin{align}
     \lambda_4 = \Delta T_\ir -  \frac{k^4 \aIR^2}{4(4+\epsb)} < - \frac{\aIR^2 k^4 (4 + \epsb)}{16 \epsb} \, .
     \label{eq:maxdetuning}
\end{align}

In order to implement a similar procedure to what was done in the previous section, we fix $\lambda_4$ first and cancel the full c.c. with an appropriate UV detuning. We start with IR brane tension:
\begin{align}
    \Delta T_\ir = \lambda_4 + \frac{k^4 \aIR^2}{4(4+\epsb)} .
\end{align}
That is we start with the initial tension to be an amount $\lambda_4$ above the minimum required value, before perturbing this solution by an additional amount $\delta T_\ir$. Solving the IR boundary condition for $A'$, we find that the critical radius, $r_c$, satisfies
\begin{align} \label{eq:rc_linear}
    e^{\epsb k \pi r_c} =\frac{\aIR k^{3/2} (4+\epsb)(\aUV + \epsb ) }{4 \aUV \vUV\epsb  (2 + \epsb)}
    \left(1-\sqrt{1+  \frac{16 \epsb \, \lambda_4}{\aIR^2 k^4 (4+\epsb)}}\right) \, ,
\end{align}
where we note that the first factor is positive as both $\aIR<0$ and $\epsb<0$. There is an upper bound on the tension, equation~\eqref{eq:maxdetuning}, which comes from the requirement that the square root is positive.

To tune away the cosmological constant for our initial solution the zeroth order tuning $\Delta T_\uv^0$ is given by equation~\eqref{eq:UVcrit_tuning}. The subleading contribution to the tension, $\Delta T_\uv^{(4)}$, is now given by
\begin{align}
    \Delta T_\uv^{(4)} &= e^{-4k\pi r_c} \left[
    \frac{- \epsb \, \lambda_4}{2 + \epsb}
     +\frac{\aIR^2 k^4 }{4 (2 + \epsb)}\left( -1 +   \sqrt{1 + \frac{16 \epsb \, \lambda_4}{\aIR^2 k^4 (4 + \epsb)}} \right) \right] \, .
     \label{eq:UVdetune4_linear}
\end{align}
As $\lambda_\uv$ is unchanged from above, the boundary value for the GW field on the UV brane $\Phi(0)$ is as in equation~\eqref{eq:GWboundaries}, but the value in the IR becomes
\begin{align}
    &\Phi(\pi) = \frac{\aIR k^{3/2}}{2\epsb} \left( \frac{4}{4 + \epsb} - 
    \sqrt{1+\frac{16\epsb \, \lambda_4}{ \aIR^2 k^4(4 + \epsb)}}    \right) \, .
\end{align}

We now perturb the tension by an amount $\delta T_\ir$ to study how the system responds.  This corresponds to a choice for the full IR brane tension:
\begin{align}
    \Delta T_\ir = \lambda_4 + \frac{k^4 \aIR^2}{4(4+\epsb)} + \delta T_\ir \, .
\end{align}
After perturbing the brane tension by $\delta T_\ir$ we again find that the cosmological constant shifts in the same way as for the quadratic case with $\epsilon>0$
\begin{equation}
    \delta \Lambda = \frac{k \kappa^2}{3} e^{-4k \pi \rc}  \delta T_{\ir} \, .
\end{equation}
In particular, the shift is given purely by the warped energy density and is insensitive to the parameters of the stabilization mechanism. To leading order in $\epsilon$ the critical radius shifts according to 
\begin{align} \label{eq:deltarcIRlin}
    \delta r_c  = \frac{ \delta T_\ir}
    {\pi  k \lambda_4 \epsilon} \, .
\end{align}
As was the case for the quadratic model with $\epsilon>0$ the fractional shift in $r_c$ can be large due to the factors of $\epsilon$ in the denominator. 

The leading order shifts in the GW field and it's derivatives are:
\begin{equation}
\begin{aligned}
    &\delta \Phi(0) = \frac{  \delta T_{\ir} }{2 k \aUV  \vUV} e^{-4 k \pi r_c} \,, 
    &&\delta \Phi(\pi) =  - \frac{\delta T_{\ir}}{\aIR k^{5/2}} + \mathcal{O(\epsilon)} \, , 
    \\
    &\delta \Phi'(0) = \frac{r_c\delta T_{\ir} }{\vUV} e^{-4 k \pi r_c}  \,, 
    &&\delta \Phi'(\pi) =  0\, .
\end{aligned}
\end{equation}
The corrections to the energy densities on each brane from the GW potentials, as defined in equation~\eqref{eq:GWpotentialsUV}, are:
\begin{equation}
\begin{aligned}
    \delta \lambda_\uv (\Phi) &= - 
    \frac{\epsilon \delta T_\ir e^{-4k \pi r_c} }{\aUV}  + \mathcal{O}(\epsilon^2)\, ,
    \\
    \delta \lambda_\ir (\Phi) &= - \delta T_\ir + \mathcal{O}(\epsilon) \, ,
\end{aligned}
\end{equation}
where again both $\delta \lambda_\uv$ and $\delta \lambda_\ir$ vanish in the limit of large $\aUV\, , \aIR$.

\section{Radion Effective Potential}
\label{sec:effectivepotential}

So far we have presented the full analysis in the higher-dimensional theory. But for low-energy physics we should be able to obtain the same results from an effective theory of the radion. Previous papers have considered such a theory but none have done the complete model allowing for a cosmological constant and the full dependence on the stabilizing potential. 

In this section we assume the radion is the lightest degree of freedom (prior to including the SM), so the effective theory is valid below the KK scale $m_{KK} = k e^{-k\pi r_c}$. The effective potential for the radion comes from integrating the bulk action for $\Phi$ coupled to gravity and combining it with the brane-localized contributions. Here we use the equations of motion to reduce the action to a sum of boundary terms plus $\Lambda$ integrated over the bulk space. Since we have four parameters $\left(\Phi(0), \, \Phi(\pi), \, \Lambda, \, r_c \right)$ and four boundary conditions, to derive the radion potential we use the boundary conditions to solve for all of the parameters except for $r_c$ before performing the dimensional reduction. Note that in doing so we are capturing the dependence of the fields and the cosmological constant on the radion.

The starting point is the bulk action~\eqref{eq:5daction}, with metric~\eqref{eq:metricansatz}:
\begin{align}
    S_{bulk} &= \int d^5x \sqrt{-g^{(5)}}  \left[ \frac{1}{2 \kappa^2} R^{(5)} - \frac{1}{2r_c^2}\Phi'^2 - V(\Phi)  \right] \,.
    \label{eq:bulkaction}
\end{align}
Here we note that the integration in equation~\eqref{eq:bulkaction} is over the full $S_1$, or two copies of the orbifold $S_1/\mathbb{Z}_2$.  Substituting the equations of motion and $\sqrt{-g_5} = e^{4A} r_c \sqrt{-g}$ we find the contribution from the bulk Lagrangian to the 4d effective potential (for $\varphi \equiv e^{-k \pi r_c}$):
\begin{align}
    V_{bulk} (\varphi) &= \sqrt{-g} \left( \frac{4A'(0)}{r \kappa^2} -  \frac{4A'(\pi)}{r \kappa^2} \varphi^4\right) 
    +
    2 r_c  \int_0^\pi d \phi \sqrt{-g} \left[
    \frac{3r_c }{\kappa^2} \Lambda e^{2A}
    - \frac{1}{r_c \kappa^2} \frac{d}{dr} \left( A' e^{4A} \right) \right]
    \, ,
    \label{eq:bulkpotential}
\end{align}
where $A$, unlike $\varphi$, is a function of the coordinate $\phi$.
The first term in equation~\eqref{eq:bulkpotential} comes from treating the singular piece in $A''$ at the boundaries, or can be equivalently derived from the Gibbons-Hawking boundary terms~\cite{Bellazzini:2013fga}. In both the UV and IR these terms combine with the total derivative term and the brane localized potentials to give two contributions to the potential that depend on the boundary values of the GW field and the warp factor:
\begin{align}
    V_{\uv} (\varphi) &=  \lambda_{\uv}\left( \Phi(0) \right) + \frac{6}{r_c \kappa^2}  A'(0)  \, ,
    \label{eq:effpotUV}
    \\
    V_{\ir} (\varphi) &= \varphi^4 \left( \lambda_{\ir}\left( \Phi(\pi) \right) - 
    \frac{6}{r_c \kappa^2} A'(\pi) \right)  \, .
    \label{eq:effpotIR}
\end{align}
These terms are both proportional to the boundary condition satisfied by the warp factor, so they vanish at the minimum of the potential. 

The remaining bulk contribution comes from the 4d Ricci scalar ($\propto \Lambda$) and is 
\begin{align}
    V_{\Lambda} (\varphi) &= \frac{3 \Lambda}{k \kappa^2} \left( 1 - \varphi^2 \right) \, .
    \label{eq:effpotEH}
\end{align}
The full effective potential is then a sum of  the three terms
\begin{align}
    V_{\eff} (\varphi) =  V_{\uv} (\varphi) + V_{\ir} (\varphi) + V_{\Lambda} (\varphi) \, .
\end{align}
We note that the $V_\Lambda$ term has often been neglected in the literature on the RS effective potential. This extra contribution to the potential explains how there can be a constant energy density in the effective theory despite the fact that the boundary potentials $V_\ir$ and $V_\uv$ both vanish on-shell. For a de-Sitter slicing this extra contribution yields a tachyonic mass term which can in principle destabilize the solution (as an example see ref.~\cite{Mishra:2022fic}). 

In section~\ref{subsec:model_potentials} we will derive the effective potentials corresponding to the two stabilizing potentials we have considered. These will be messy so to highlight the features we expect, in the following subsection we first present a toy model that captures many of the essential features of the full potential. The model highlights why the low-energy cosmological constant should be unsuppressed by the parameters of the stabilization mechanism when energy is added to the system, and why a superficial analysis would suggest otherwise.

\subsection{Simplified Model}

In this section we discuss a simplified model for the radion potential. The model illustrates how the potential na\"ively leads to a suppressed c.c. when we perturb by adding energy, whereas the true c.c. is unsuppressed once the induced shift in radion vev is included. This reproduces the results of section~\ref{sec:5dmodel} where the c.c. resulting from detuned brane tensions was unsuppressed by the parameters of the stabilizing potential. Here our simplified potential is given by the general potential~\eqref{eq:Veff_general}, after dropping the $\varphi^{4+2\epsilon}$ term and setting $\lambda_{4+\epsilon} =1$.

The potential we consider is then
\begin{align}
    V(\phi) = \phi^4 \left(\phi^\epsilon +T \right)\, ,
    \label{eq:toypotential}
\end{align}
where $\epsilon$ is a small parameter. The minimum of $V$ is at $\phi = \phi_c$ with
\begin{align}
    \phi_c = \left( -\frac{T}{1+ \frac{\epsilon}{4}} \right)^{1/\epsilon}\, ,
\end{align}
and a cosmological constant
\begin{align}
    \Lambda_0 = V(\phi_c) = \frac{\epsilon}{4+\epsilon} \phi_c^4 T  \, .
    \label{eq:toycc}
\end{align}
This cosmological constant is the analog of the $\mathcal{O}(\varphi^4)$ contribution that needed to be cancelled by shifting the UV tension, as was done in equations~\eqref{eq:UVcrit_tuning4} and~\eqref{eq:UVdetune4_linear}.

In response to a shift $T \to T + \delta T$, it appears that $\Lambda $ shifts by an amount $\propto \epsilon \delta T$. However, this does not take into account the  change in $\phi_c$. The new minimum after the perturbation is
\begin{align}
    \tilde \phi_c = \left( -\frac{T + \delta T}{1+ \frac{\epsilon}{4}} \right)^{1/\epsilon} \, ,
\end{align}
where the $\phi$ field has shifted by an amount
\begin{align}
    \frac{\delta \phi_c}{\phi_c} = \frac{\delta T}{\epsilon T}  + \mathcal{O}( \delta T^2 ) \,.
\end{align}
The resulting shift in the cosmological constant is
\begin{align}
   \tilde \Lambda &=  \frac{\epsilon}{4+\epsilon} \tilde \phi_c^4 (T + \delta T) \, ,
   \\
   &= \frac{\epsilon}{4+\epsilon} \phi_c^4 T + \frac{\epsilon}{4+\epsilon} \phi_c^4 \delta T
   +\frac{4\epsilon}{4+\epsilon} \phi_c^3 \delta \phi_c  T 
   + \mathcal{O}( \delta T^2 ) \, ,
   \\   
   &= \frac{\epsilon}{4+\epsilon}\phi_c^4 T 
   +  \phi_c^4 \delta T  + \mathcal{O}( \delta T^2 ) \, ,
\end{align}
so the leading correction to $\Lambda$ is
\begin{align}
    \delta \Lambda = \phi_c^4 \delta T \, .
\end{align}
The fact that $\phi$ shifts by $\mathcal{O}(1/\epsilon)$ compensates for any would-be $\epsilon$ suppression. Taking into account this shift, the response of the cosmological constant to a change in the energy density is unsuppressed by the parameter $\epsilon$. Given that the effective potential reduces to terms that vanish when the boundary conditions are satisfied (equations~\eqref{eq:effpotUV} and~\eqref{eq:effpotIR}), the cosmological constant in the 4d theory is simply the energy density put in, without any suppression from parameters of the stabilization model.

In the remainder of the section we will derive the analogous result for the potential derived from the Goldberger-Wise models we discussed in the previous section. We will see that the full effective potential reproduces the results of section~\ref{sec:5dmodel} and leads to unsuppressed vacuum energies, exactly reproducing the results of the previous sections.

\subsection{Potentials from different models}

\label{subsec:model_potentials}

The 5d system can be characterized by four quantities:  the values of the GW field on the boundaries, $\Lambda$, and $r_c$. To derive the effective potential we use three of the boundary conditions to fix $\Lambda$, $\Phi(0)$~and~$\Phi(\pi)$, leaving $r_c$ undetermined. We initially focus on the model with quadratic boundary potentials on both branes, and take the nonzero cosmological constant to arise from an IR detuning. We give expressions for the quadratic/linear model of section~\ref{subsec:linear} at the end of the section.

Our first step is to solve the boundary condition for $A'(0)$. 
This gives an expression for $\Lambda$ in terms of the model parameters:
\begin{align}
    \frac{3 \Lambda}{k \kappa^2} = \Delta T_\uv + \frac{k v_\uv^2 \aUV \epsb}{\aUV + \epsb} + \varphi^4 f_{\Lambda} (\varphi^\epsb) + \mathcal{O} (\varphi^8) \, ,
    \label{eq:ccgeneral_1}
\end{align}
where $f_\Lambda$ is a polynomial of degree two and we have written the expression this way to encapsulate the model-dependence in $f$. We will introduce a detuning on the IR brane, so we tune $\Delta T_\uv$ as in equation~\eqref{eq:UVcrit_tuning} to get
\begin{align}
    \frac{3 \Lambda}{k \kappa^2} = \Delta T^{(4)}_\uv + \varphi^4 f_{\Lambda} (\varphi^\epsb) + \mathcal{O} (\varphi^8) \, ,
    \label{eq:ccgeneral}
\end{align}
where ultimately we will need to choose $\Delta T^{(4)}_\uv$ to be the critical value in equation~\eqref{eq:UVcrit_tuning4} to set the cosmological constant to zero at the initial minimum. This way of writing the cosmological constant incorporates the dependence of $\Lambda$ on the stabilization mechanism through $f_\Lambda$  and the critical brane tension, which we shortly give for particular models. The quantity $\Delta T^{(4)}_\uv$ is a parameter of order $\mathcal{O}(\varphi^4)$, so we drop terms involving $\Delta T^{(4)}_\uv$ multiplying powers of $\varphi$ when we truncate the potential to $\mathcal{O}(\varphi^4)$ below.

As we have solved the boundary condition for $A'(0)$ to get equation~\eqref{eq:ccgeneral_1}, the UV boundary term $V_\uv$ does not contribute. So when we substitute the above expression for $\Lambda$ into the contribution from the IR brane potential, the full effective potential for the radion has the general form
\begin{align}
    V_{\eff}  = \varphi^4 \left( f_{\ir} (\varphi^\epsb) + f_{\Lambda} (\varphi^\epsb) \right) + \Delta T^{(4)}_\uv \, ,
    \label{eq:Veff_general2}
\end{align}
where $f_\ir$ is a quadratic polynomial that depends on the specifics of the stabilization mechanism. This potential is minimized for 
\begin{align}
     \epsb \varphi^\epsb \left( f'_\Lambda + f'_\ir \right) + 4 \left( f_\Lambda + f_\ir\right) = 0 \, .
     \label{eq:radmin}
\end{align}
We also know that the IR boundary condition must be satisfied  for the solution, $\varphi_0$, to this equation, i.e.
\begin{align}
    f_\ir (\varphi_0) = 0 \, .
\end{align}
The cosmological constant can then be set to zero by fixing 
\begin{align}
    \Delta T^{(4)}_\uv = - f_\Lambda (\varphi^\epsb_0) \varphi_0^4 \, ,
    \label{eq:crit4}
\end{align}
as was done in equations~\eqref{eq:UVcrit_tuning4} and~\eqref{eq:UVdetune4_linear} for the two models we considered.

We find the shift in $\Lambda$ due to a change in $\delta T_\ir$, which is given by equation~\eqref{eq:DLambdaIR}, namely $ \delta \Lambda = \frac{k \kappa^2}{3} e^{-4k \pi \rc}  \delta T_{\ir} $, for this form of the potential. The potential evaluated at $\varphi_1 = \varphi_0 + \delta \varphi$ is
\begin{align}
    V_{\eff} (\varphi_1)  &= \Delta T^{(4)}_\uv + \varphi_1^4 \left( f_{\ir} (\varphi_1^\epsb) + f_{\Lambda} (\varphi_1^\epsb) \right) + \delta T_\ir \varphi_1^4 \, .
\end{align}
Treating $\delta \varphi$ and $\delta T_\ir$ as perturbations this becomes:
\begin{equation}
\begin{aligned}
    V_{\eff} (\varphi_1)  &\simeq \Delta T^{(4)}_\uv + \varphi_0^4 \left( f_{\ir} (\varphi_0^\epsb) + f_{\Lambda} (\varphi_0^\epsb) \right) + \delta T_\ir \varphi_0^4 
    \\
    &= \delta T_\ir \varphi_0^4 \, .
\end{aligned}
\end{equation}
The cancellation between terms comes from using equations~\eqref{eq:radmin} and~\eqref{eq:crit4}. That is, the result comes from the fact that we are expanding around a point where both $V$ and $V'$ are zero, so many of the terms drop out. This result assumed nothing about the functions $f_\Lambda$ and $f_\ir$, so is independent of the details of the stabilization mechanism.
\\

For the quadratic/quadratic model considered in section~\ref{subsec:quadratic} these functions are
\begin{align}
    f_{\Lambda} (\varphi^\epsb) &= k \vUV \epsb  (2 + \epsb) \varphi^\epsb 
    \left( \vIR  - \varphi^\epsb \vUV \right)
    \, ,
    \nonumber
    \\
    f_{\ir} (\varphi^\epsb) &= k \vIR^2 (4+\epsb) \left( 1
    - \frac{\vUV \left(2 + \epsb \right)}{ \vIR} \varphi^\epsb 
    +\frac{ \vUV^2 (2 + \epsb)^2}{\vIR^2 (4+\epsb)}
    \varphi^{2\epsb} \right)
    \, ,
    \label{eq:Veff_functions}
    \\
    f_{\Lambda} (\varphi^\epsb) + f_{\ir} (\varphi^\epsb) &= k\vIR^2 (4+\epsb)  
    - 4 k \vUV \vIR (2 + \epsb)\varphi^\epsb 
    + 2 k \vUV^2 (2 + \epsb) \varphi^{2\epsb} 
    \, ,
    \nonumber
\end{align}
where for clarity we have taken the large-$\alpha$ limit.
It is easily checked that substituting these functions into~\eqref{eq:radmin} to find the minimum $\varphi_0$ reproduces the solution of equation~\eqref{eq:minimum1}. Furthermore, at the minimum $f_\ir (\varphi_0^\epsb) = 0$ and the IR boundary condition is satisfied, as expected. The functions $f_\Lambda, f_\ir$ for the model with a linear boundary potential in the IR are (taking $\aUV \to \infty$ but keeping finite $\aIR$)
\begin{align}
    f_{\Lambda} (\varphi^\epsb) &= 
    \frac{k \vUV \epsb (2 + \epsb)}{(4 + \epsb)} \varphi^\epsb\left( 
    \vUV  \epsb \varphi^\epsb - \frac{\aIR k^{3/2}}{2} \right)
    \, ,
    \nonumber
    \\
    f_{\ir} (\varphi^\epsb) &=  \lambda_4
    + \frac{k(2 + \epsb) \vUV}{2} \varphi^\epsb \left( 
    \aIR  k^{3/2} - \varphi^\epsb \frac{2 \ \vUV \epsb  (2 + \epsb)}{4 + \epsb} \right)
    \, ,
    \\
    f_{\Lambda} (\varphi^\epsb) + f_{\ir} (\varphi^\epsb) &= 
    \lambda_4
    + \frac{\vUV (4 + 2\epsb)}{4 + \epsb} \varphi^\epsb 
    \left( \aIR  k^{3/2}- \vUV \epsb  \varphi^\epsb\right)
    \, .
     \nonumber
\end{align}
So the effective low-energy potential is equation~\eqref{eq:Veff_general2} with the expressions as above. This takes the expected form but with model-dependent parameters derived from the five-dimensional potential. Again minimizing the potential reproduces the result for $\varphi^\epsb$ in equation~\eqref{eq:rc_linear}. Setting the UV detuning to the critical value in~\eqref{eq:UVdetune4_linear} leads to a vanishing $\Lambda$, and perturbing the brane tension by an amount $\delta T_\ir$ reproduces the shifts in $\Lambda$ and $r_c$ found in section~\ref{subsec:linear}.

\subsection{Radion Mass}
\label{subsec:radionmass}

With the effective potential at hand we now compute the radion mass, including finite-$\alpha$ corrections. For quadratic potentials in both the IR and UV the expression is
\begin{align}
   m_\varphi^2 = 4 k^{-2}e^{-2k \pi r_c} \epsb^{1/2} \left( 2 k \vIR^2 \epsb  - \delta T_{\ir} \right)\sqrt{ \frac{\aIR }{\aIR+4}} \, ,
   \label{eq:radmassquad}
\end{align}
expanding to $\mathcal{O}(\delta T_\ir)$ and keeping the leading $\epsb$ dependence. The brane tension must satisfy $\Delta T_\ir < k \epsb v_\ir^2 $ to have a positive squared radion mass, the same bound as was found in~\eqref{eq:TIRbound1} to not destabilize the solution. For the holographic model the mass is
\begin{align}
   m_\varphi^2 = -2 \epsb k^{-2}e^{-2k \pi r_c} \left( \lambda_4 + \delta T_\ir \right) \, .
   \label{eq:radmasslin}
\end{align}
In this case the lower bound on the detuning of the IR brane tension~\eqref{eq:mindetuning} corresponds to $\lambda_4 + \delta T_\ir >0$, so any brane tension not satisfying this bound again leads to a negative $ m_\varphi^2$, highlighting the instability at this point.

In both the models we considered, the shift in $r_c$ from detunings on the different branes are parametrically given by 
\begin{align}
    &\delta r_c \sim \frac{\delta T_\ir e^{- 4k \pi r_c}}{k \pi m_\varphi^2 m_{KK}^2} \, ,
    &&\delta r_c \sim \frac{\delta T_\uv e^{-2 k \pi r_c}}{k \pi m_\varphi^2 m_{KK}^2} \, .
    \label{eq:shifts-general}
\end{align}
Written in this form makes clear the origin of the large potential shifts in $r_c$ (enhanced by $1/\epsb$) found in equations~\eqref{eq:deltarcquad} and~\eqref{eq:deltarcIRlin}, namely that the radion mass can be parametrically suppressed relative to $m_{KK}$ by the conformal breaking parameter $\epsb$.

We also note that in a model with the full IR potential we expect a radion-higgs coupling that generates an irreducible contribution to the radion mass. The radion-higgs coupling in the effective theory is 
\begin{align}
    V(H, \varphi) = \frac{\lambda}{4} \left(H^\dag H - v_H^2 \varphi^2\right)^2 \, , 
\end{align}
where $v_H$ is a UV parameter related to the electroweak scale by $v_\ew = v_H \langle \varphi \rangle$. The electroweak phase transition will then cause a change in $r_c$ of order
\begin{align}
    \delta r_c \sim \frac{\lambda v_\ew^4}{m_\varphi^2 m_{KK}^2} \, .
\end{align}
This contribution is suppressed by the ratio of the electroweak scale $v_\ew$ to the KK scale $m_{KK} = ke^{-k\pi r_c}$, which must be at least $v_\ew / m_{KK} \lesssim .1$ from collider bounds~\cite{CMS:2018dqv,CMS:2018ipl,ATLAS:2021uiz, CMS:2021ctt}, but can be significant in scenarios when $\epsb \ll 1$ and the radion mass is extremely small.

\section{Supersymmetry Breaking and the Effective Theory}

\label{sec:SUSY}

As an important application of the above analysis, having seen how the space adjusts in the presence of perturbing energy, we can also better 
understand how supersymmetry (SUSY) breaking can be communicated from one brane to another. 
SUSY breaking in higher-dimensional space arises in many contexts, most notably in string theory where supersymmetry breaking is a prerequisite for a nonzero cosmological constant. KKLT, for example, introduces an antibrane into a warped throat to generate SUSY breaking energy at an intermediate scale that can compensate for an AdS energy generated by fluxes~\cite{Kachru:2003aw}.

Several papers have addressed the issue of how SUSY-breaking might be communicated in this context based on a 4d effective field theory~\cite{Choi:2005ge,Choi:2005uz,Randall:2019ent}. This however introduces a puzzle about how SUSY breaking can be communicated from the IR to fields on the UV brane without violating sequestering, since the warped space is by assumption a space in which every 4d slice has the same non-vanishing 4d c.c.

The key insight (and puzzle) is that we have found a consistent solution without ever mentioning $\sqrt{\delta T}$, which is the usual supersymmetry-breaking order parameter. We have seen that the space adjusts through the change in the stabilizattion radius, which is not itself a supersymmetry-breaking order parameter, although it can induce non-zero supersymmetry-breaking auxiliary terms. We emphasize this shift is sensitive to the change in energy, and hence changes only in response to $\delta T$.

In this section we give general arguments why the shift in the stabilizing fields scales as $\delta T$ also in supersymmetric models. Because this is much smaller than the usual SUSY breaking order parameter, $\sqrt{\delta T}$, the rearrangement of the geometry  in response to a perturbation does not violate sequestering. This allows for the dominant communication of SUSY breaking to come from the compensator $F$-term,  which does scale as $\sqrt{\delta T}$. These findings are consistent with the results of refs.~\cite{Luty:1999cz, Luty:2000ec, Son:2008mk}, which considered supersymmetric models inspired by RS-like setups.

As shown in section~\ref{sec:effectivepotential} the effective potential reduces to a sum of boundary terms and a contribution to the 4d cosmological constant. In a supersymmetric model this means that the effective potential at the minimum takes the form:
\begin{align}
    V_{\rm eff} &=   \frac{3 \Lambda}{k \kappa^2} \left( 1 - \varphi^2 \right) \, ,
    \nonumber
    \\
     \frac{3 \Lambda}{k \kappa^2} &= \varphi^4 \left( \sum |F_\ir (\varphi)|^2 \right)
    + \sum |F_\uv (\varphi)|^2\, .
    \label{eq:SUSYpotential}
\end{align}
where $F_\ir, F_\uv$ are the $F$-terms for UV and IR localized contributions, respectively, which will in general be $\varphi$-dependent. Note that the supersymmetric solution with no $F$-terms corresponds to  tuned tensions on both branes, as discussed in ref.'s~\cite{Bergshoeff:2000zn, Fujita:2001bd, PaccettiCorreia:2004suu}. To simplify the discussion we define
\begin{align}
    &\mathcal{E}_\ir (\varphi)= \sum |F_\ir (\varphi)|^2 \, ,
    &&\mathcal{E}_\uv (\varphi)= \sum |F_\uv (\varphi)|^2 \, ,
\end{align}
where both $\mathcal{E}_\ir$ and $\mathcal{E}_\uv$ are strictly positive.

We then ask how the radion shifts in the presence of this SUSY-breaking vacuum energy. Our starting point is a supersymmetric solution with $\Lambda = 0$, which is minimised at $\varphi_0$. This implies the conditions 
\begin{align}
    &\mathcal{E}_\ir (\varphi_0) =  \mathcal{E}_\uv (\varphi_0)  = 0 \, ,
    && \varphi_0^4 \mathcal{E}'_\ir (\varphi_0) +  \mathcal{E}'_\uv (\varphi_0)  = 0 \, .
\end{align}
For simplicity in what follows we will take $\mathcal{E}_\uv$ to vanish identically, but the analysis can be adapted to the case where the leading term in $\mathcal{E}_\uv$ is $\mathcal{O}(\varphi^4)$. The mass of the radion in this case is given by
\begin{align}
    k^2 m_\varphi^2 = \frac{\partial^2 V}{\partial \varphi^2 } \bigg |_{\varphi_0}= \varphi_0^4  \mathcal{E}''_\ir (\varphi_0) \, .
\end{align}

If SUSY breaking comes from IR-localized fields then (at least) one of the $F_\ir$-terms gets a vev, which we take to be constant. This can be modeled by shifting $\mathcal{E}_\ir \to \mathcal{E}_\ir + \delta T_\ir$. As $\Lambda \sim \delta T_\ir \varphi_0^4$, the $\Lambda \varphi^2$ term in the effective potential is negligible and the new minimum is at $\varphi$ such that
\begin{align}
    4 \varphi^3 \left( \delta T_\ir + \mathcal{E}_\ir (\varphi) \right)+ \varphi^4 \mathcal{E}'_\ir (\varphi)  = 0 \, .
\end{align} 
If we perturb around the previous solution $\varphi = \varphi_0 +\delta \varphi$ then we have
\begin{align}
    4 \varphi_0^3 \delta T_\ir + \varphi_0^4 \mathcal{E}''_\ir (\varphi_0) \delta \varphi = 0 \, .
\end{align}
The second term is just $ k^2 m_\varphi^2 \, \delta \varphi $, so this gives: 
\begin{align}
    \delta \varphi = - \frac{4 \varphi_0^3 \delta T_\ir}{ k^2 m_\varphi^2}
\end{align}
This corresponds to a shift in $\delta r_c$ 
\begin{align}
    \delta r_c = -\frac{1}{k \pi} \frac{\delta \varphi}{\varphi_0} = \frac{4 \delta T_\ir \varphi_0^4}{k \pi m_{KK}^2 m_\varphi^2}\, ,
\end{align} 
where we have identified $m_{KK} = k \varphi_0$.

If the SUSY-breaking vacuum energy is on the UV brane instead, then we set $\mathcal{E}_\uv = \delta T_\uv = const$. This introduces a quadratic term in $\varphi$ into the potential. Minimizing the potential in this case gives:
\begin{align}
   4\varphi^3 \mathcal{E}_\ir + \varphi^4 \mathcal{E}_\ir '- 2 \delta T_\uv \varphi = 0 \, .
\end{align}
Note that this implies that some of the $F_\ir$ must also turn on in order for the potential to remain stabilized. Repeating an analysis similar to above, we can derive the shift in $r_c$ in this case as 
\begin{align}
    \delta r_c \sim \frac{ \delta T_\uv  \varphi_0^2}{k \pi m_{KK}^2 m_\varphi^2}  \, .
 \end{align}

In both cases this matches what was found in sections~\ref{sec:5dmodel}~\&~\ref{sec:effectivepotential} (see equation~\eqref{eq:shifts-general}) and shows that there is always a $\mathcal{O}(\delta T)$ shift in the fields due to the fact that $r_c$ changes at that order. This is independent of how SUSY-breaking is communicated or whether the energy is perturbed due to supersymmetry breaking. Note that this corresponds only to the shifts in fields needed to provide the stable background. We expect the dominant contribution to supersymmetry breaking mass parameters arises from the $F$ terms of fields involved in supersymmetry breaking or from anomaly mediation, which scale as $\sqrt{\delta T}$, although the results can be more complicated as we show in a companion paper \cite{Nee:2025nmi}. Because the original shift required to maintain the warped geometry is $\mathcal{O}(\delta T)$, it is naturally consistent with sequestering.

Ref.~\cite{Son:2008mk} also considered a supersymmetric higher-dimensional model and found that the stabilization mechanism itself  preserves supersymmetry only for a specific value of $r$. This is easily seen as the $F$-term equations for $\Phi$ are first order, so vanish only when the coefficients of one of the exponentials in equation~\eqref{eq:Phisol} vanishes. In our case the SUSY-preserving solution corresponds to setting $C_1=0$ in~\eqref{eq:Phisol}. Any shift in $r_c$ from dynamics on either brane will then lead to $C_1$ turning on at $\mathcal{O}(\delta r_c)$ and breaking supersymmetry in the bulk, with a corresponding $F$ term of order $\delta T$. A detailed analysis of these effects is presented in a separate paper~\cite{Nee:2025nmi}.

\section{Conclusions}

In this paper we revisited the construction of effective theories coming from warped extra dimensions. Our analysis focused on the RS model with two branes stabilized via various implementations of the Goldberger-Wise mechanism. The main question we have answered is how the stabilized solution changes when energy is added locally to either of the branes. 
To do so, we  kept the leading backreaction terms from $\Phi$ and $\Lambda$ as well as the corrections due to finite couplings on the branes. This allowed us to solve the complete model in the 5d picture before constructing the 4d effective potential.  Working in the 5d setup we found how $\Phi$, the warp factor $A$ and the c.c. $\Lambda$ change in response to a change in energy density on either brane. 

We then constructed the 4d effective potential, and used it to reproduce the results found in the 5d theory. In order to derive $\Lambda$ from the 4d theory it was necessary to include the contributions to the effective potential from the Einstein-Hilbert term, as the other terms are proportional to boundary terms and vanish on-shell. The effective potential also allowed us to compute the radion mass, which we found to be suppressed by the mass of $\Phi$ for our simple GW-inspired models. With our analysis we could also classify which combinations of boundary potentials and bulk squared masses for $\Phi$ led to stabilized models.

Our main results are in Sections~\ref{sec:5dmodel} and~\ref{sec:effectivepotential}, and
incorporate a number of effects that are either incomplete or missing from many similar analyses in the literature, namely:
\begin{itemize}
    \item We include finite-$\alpha$ corrections  to the GW mechanism. For finite $\alpha$, the boundary condition is not pure Dirichlet but reproduces the Dirichlet boundary condition as $\alpha \to \infty$.
    \item We include the backreaction of the GW field on the metric to leading order, allowing us to get the full radion dependence of the low-energy potential
    \item We self-consistently include the effect of a non-vanishing c.c. This includes the backreaction of $\Lambda$ on the warp factor, which is necessary to keep in order to solve the boundary condition for $A$. This also allows us to see explicitly the origin of an unsuppressed 4d c.c.\ in the presence of an IR perturbation. 
    \item We include the cosmological constant term $V_\Lambda$ (see eq.~\eqref{eq:effpotEH}) in the effective potential. Including this term is necessary to satisfy the boundary conditions  at the minimum of the potential.
\end{itemize}

We found that after adding energy density to either of the branes, the warp factor and GW field adjust so that each slice of the extra dimension has the same cosmological constant, but the back-reacted warp factor differs appropriately from the vacuum solution with the same 4d c.c. The energy density, which was added locally, is therefore reflected throughout the extra dimension.  When the GW potential strongly fixes the fields on the boundaries (that is, large $\alpha$), the energy on the other brane does not significantly change. Here it is the change in warp factor that is primarily responsible for maintaining a constant 4d c.c. throughout the bulk. The cosmological constant is  unsuppressed by the parameters of the stabilization, which is contrary to what was found in scalar gravity models without a UV brane, for example, and also contradicts some claims that have been made in the literature.

There is one further intriguing aspect to our result when interpreted from a holographic perspective, which could potentially shed insight into the UV cancellation of the four-dimensional cosmological constant. The issue of the small c.c.\ is puzzling, but is yet more puzzling due to the contributions from the IR. For example, though the electroweak and QCD phase transitions involve energies far greater than the c.c.\ itself, those energies are nonetheless much smaller than the UV scale.

In our paper we did not construct an explicit CFT dual for our 5d theory. But were
one to exist, the GW field would have a CFT interpretation. In the dual interpretation,
the modification of the warp factor required for consistent jump conditions would imply
consistent scaling of the 4d field theory so that the net cosmological constant is always
determined by the sum of the UV and warped IR contributions, even though the boundary
condition in the UV is locally determined only by the brane energy.  In the gravitational theory, an adjustment in the IR is directly communicated to the UV via a perturbation in the stabilizing field. On the CFT side, this corresponds to the required change in UV boundary condition to lead to the  modified IR theory. In the holographic theory, the UV and IR are connected by the CFT scaling, which  in 5d is encoded in  the shift in the bulk scalar field.

Finally, we briefly explored the implications if the theory is supersymmetric. We argued that the 5d theory remains sequestered even though the stabilization mechanism transmits any SUSY-breaking energy throughout the bulk space. We leave a more complete analysis of the supersymmetric model for a separate work~\cite{Nee:2025nmi}.

In summary, the effective theory of the radion is indeed valid when the radion is parameterically the lightest field. However, the analysis can be more subtle than is generally assumed, with potentially important and measurable low-energy consequences. This should be further studied in future work.

\section*{Acknowledgements}

We thank Brando Bellazzini, Csaba Csaki, Raphael Flauger, Rashmish Mishra, Michele Papucci, Alex Pomarol, Riccardo Rattazzi, Martin Schmaltz, Raman Sundrum, John Terning, and Neil Weiner for useful discussions. We would like to acknowledge GRASP Initiative funding provided by Harvard University. The work of SL is in part supported by the NSF grant PHY-1915071. The work of MN is supported by NSF Award PHY-2310717. 
The work of LR has been supported by NSF grants PHY-1620806 and PHY-1915071, the Chau Foundation HS Chau postdoc award, the Kavli Foundation grant “Kavli Dream Team,” and the Moore Foundation Award 8342.
We thank NYU, UC San Diego, and IPhT for their hospitality.

\appendix

\bibliographystyle{utphys}
\bibliography{refs}

\end{document}